\documentclass[12pt,reqno]{amsart}

\usepackage{amsmath,amsfonts,amsthm,cite,amssymb}

\usepackage{a4wide}
\usepackage[footnotesize,hang,sf]{caption}
\usepackage{times}
\usepackage{psfrag}
\usepackage[dvips]{graphicx}
\usepackage{pstricks}

\newcommand{\I}{\mathrm{i}}
\newcommand{\E}{\mathrm{e}}
\newcommand{\middle}{}
\usepackage{titlesec}
\titleformat{\section}
{\bfseries\scshape\centering}
{\thesection.}{.5em}{}
\titleformat{\subsection}
{\rmfamily\bfseries}
{\thesubsection}{.5em}{}
\numberwithin{equation}{section}
\newcommand{\vv}{\vec}
\newtheorem{theorem}{Theorem}
\newtheorem{prop}[theorem]{Proposition}
\newtheorem{coro}[theorem]{Corollary}

\DeclareMathOperator{\sign}{sign}

\DeclareMathOperator{\sh}{sh}
\DeclareMathOperator{\ch}{ch}

\DeclareMathOperator{\CS}{CS}

\DeclareMathOperator{\Tors}{Tors}

\begin{document}

\renewcommand{\thefootnote}{\fnsymbol{footnote}}
\baselineskip 16pt
\parskip 8pt
\sloppy




\title{On the Quantum  Invariants for the
  Spherical Seifert Manifolds}


\author{Kazuhiro \textsc{Hikami}}


  \address{Department of Physics, Graduate School of Science,
    University of Tokyo,
    Hongo 7--3--1, Bunkyo, Tokyo 113--0033,   Japan.
    }
    
    \urladdr{http://gogh.phys.s.u-tokyo.ac.jp/{\textasciitilde}hikami/}

    \email{\texttt{hikami@phys.s.u-tokyo.ac.jp}}


\date{April 24, 2005. Revised on February 4, 2006.}

\begin{abstract}
We study the
Witten--Reshetikhin--Turaev SU(2) invariant for the Seifert manifolds
$S^3/\Gamma$ where $\Gamma$ is a finite subgroup of SU(2).
We show that the WRT invariants can be written in terms of the Eichler
integral of  modular forms with half-integral weight,
and we give  an exact asymptotic expansion of the invariants
by use of the nearly modular property of the Eichler integral.
We further discuss that
those modular forms have a direct connection with the polyhedral group
by showing that
the invariant  polynomials of  modular forms
satisfy the polyhedral equations associated to $\Gamma$.

\end{abstract}





\maketitle

\section{Introduction}

Since the Witten invariant
for 3-manifold 
was introduced~\cite{EWitt89a} as
the Chern--Simons path integral,  studies of the quantum invariants
have been  much developed.
The Witten invariant
was later  redefined mathematically rigorously by
Reshetikhin and Turaev~\cite{ResheTurae91a} by use of the surgery
description of the 3-manifold and the colored Jones polynomial for links.

As was already pointed out in Witten's original paper~\cite{EWitt89a}
(see also Ref.~\citen{Atiya90Book}),
it is expected that classical topological invariants for 3-manifold
$\mathcal{M}$
could be extracted from asymptotic behavior of
the Witten--Reshetikhin--Turaev (WRT) partition function
$Z_k(\mathcal{M})$
due to that  the saddle point of the Chern--Simons path integral corresponds to
the flat connection.
Explicitly the SU(2)  WRT invariant could behave
as~\cite{FreeGomp91a,EWitt89a,Atiya90Book}
\begin{equation*}
  Z_k(\mathcal{M})
  \sim
  \frac{1}{2} \,
  \E^{- \frac{3}{4} \pi \I}
  \sum_\alpha
  \sqrt{T_\alpha(\mathcal{M})} \,
  \E^{- \frac{2 \pi \I}{4} I_\alpha} \,
  \E^{2 \pi \I  (k+2) \CS(A_\alpha) }
\end{equation*}
in large $k$ limit.
Here $T_\alpha$, $I_\alpha$, and $\CS(A_\alpha)$, respectively
denote the Reidemeister--Ray--Singer torsion, spectral flow, and the Chern--Simons
invariant.
By this observation, much attention has been paid on 
analysis  of the WRT
invariants~\cite{LCJeff92a,RLawre95a,RLawre96a,LawreRozan99a,Rozan95a,LRozansky95a,Rozan96c,Rozan96e}.


Recently it was clarified that the WRT
invariant for the Poincar{\'e} homology sphere
can be rewritten in terms of the Eichler integral of the modular form with
half-integral weight~\cite{LawrZagi99a}
(see, \emph{e.g.}, Ref.~\citen{SLang76Book} for classical definition of the
Eichler integral of modular form with integral weight).
As a consequence a
``nearly'' modular property of the Eichler integral enables us to
compute an exact asymptotic expansion of the WRT invariant.
Later it was shown~\cite{KHikami04b,KHikami04e,KHikami04f} that the
WRT invariants for the Seifert homology
spheres
can also be written in terms of the Eichler integrals of the
half-integral weight modular forms.
Based on this intriguing structure, topological invariants such as the
Reidemeister torsion, spectral flow, the Casson invariant, and the Ohtsuki
invariant,
can be  reinterpreted from the viewpoint of  modular forms.

\begin{figure}[htbp]
  \centering
  \begin{psfrags}
    \psfrag{x}{$-b$}
    \psfrag{a}{$a_1/b_1$}
    \psfrag{b}{$a_2/b_2$}
    \psfrag{c}{$a_3/b_3$}
    \includegraphics[scale=0.8, bb=-60 -60 60 72]{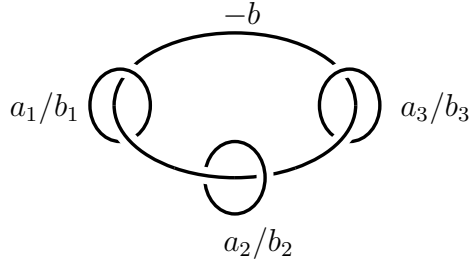}
  \end{psfrags}
  \caption{Surgery description of the Seifert manifold with three
    singular fibers
    $M(b; (a_1, b_1), (a_2, b_2) , (a_3, b_3))$.}
  \label{fig:surgery}
\end{figure}

In this article as a continuation of Ref.~\citen{KHikami04b},
we study the WRT invariant for the Seifert manifold with 3 singular
fibers
(the Brieskorn manifold),
$
\mathcal{M}=
M\left(b; (a_1, b_1), (a_2, b_2), (a_3, b_3)\right)
$~\cite{JMiln75a}.
This 3-manifold has
a surgery description as in Fig.~\ref{fig:surgery},
and
throughout this article for our convention we depict it as
\begin{equation}
  \label{Seifert_surgery}
  \raisebox{-1.6cm}{
    \begin{pspicture}(-2,-2)(2,1)
      \psdots[dotsize=4pt 2](-1,0)(0,0)(1,0)(0,-1)
      \pscustom[linewidth=1pt]{
        \psline{-}(-1,0)(1,0)
        \psline{-}(0,0)(0,-1)
      }
      \rput{0}(-1,0.5){${a_1}/{b_1}$}
      \rput{0}(0,0.5){$-b$}
      \rput{0}(1,0.5){${a_3}/{b_3}$}
      \rput{0}(0,-1.5){${a_2}/{b_2}$}
    \end{pspicture}
  }
\end{equation}
The fundamental group of the Seifert manifold $\mathcal{M}$
is written as
(see, \emph{e.g.}, Ref.~\citen{Montesi87Book}).
\begin{equation}
  \label{pi1_M}
  \pi_1 (\mathcal{M})
  =
  \left\langle
    x_1, x_2, x_3, h
    ~\middle|~
    \text{$h$ is center},
    x_i^{~a_i} = h^{-b_i} ,
    x_1 \, x_2 \, x_3 = h^{b}
  \right\rangle
\end{equation}
and it is a  homology sphere iff $a_i$ are pairwise coprime
integers.

Hereafter,
among the 3-fibered Seifert manifolds~\eqref{Seifert_surgery},
we study the spherical Seifert manifolds $S^3/\Gamma$ where $\Gamma$
is a finite subgroup of SU(2)~\cite{JMiln75a}.
We
define
$M(p_1,p_2, p_3)$
by
\begin{equation}
  {M}(p_1,p_2, p_3)
  =
  \text{SU(2)}/ \Gamma
\end{equation}
where
$\Gamma$ is a discrete subgroup classified as in
Table~\ref{tab:spherical}.
The triples $(p_1,p_2,p_3)$
are solutions of inequality
\begin{equation}
  \label{condition_spherical}
  \frac{1}{p_1} + \frac{1}{p_2} + \frac{1}{p_3} > 1
\end{equation}
with $p_j \in \mathbb{Z}_{\geq 2}$.
The order of $\Gamma$ is given by
\begin{equation*}
  4 \,
  \left(
    \frac{1}{p_1} +   \frac{1}{p_2} +   \frac{1}{p_3} -1
  \right)^{-1}
\end{equation*}
Only the manifold $M(2,3,5)$ in  Table~\ref{tab:spherical} is
homology 3-sphere, \emph{i.e.}, the Poincar{\'e} homology sphere.
Note that the manifold  $M(p_1, p_2, p_3)$ is obtained by
intersecting
the Brieskorn surface
\begin{equation}
  z_1^{~p_1} +   z_2^{~p_2} +   z_3^{~p_3} =0
  \label{Brieskorn_surface}
\end{equation}
with unit sphere
$
\left| z_1 \right|^2 + \left| z_2 \right|^2 + \left| z_3 \right|^2
=1
$.

\begin{table}[htbp]
  \centering
  \begin{equation*}
    \renewcommand{\arraystretch}{1.4}
    \begin{array}{c||c|c|c|c|}
      \mathcal{M} = M(p_1, p_2, p_3)&
      \begin{array}{c}
        \text{Seifert invariant} \\
        (b; (a_1, b_1), (a_2, b_2), (a_3, b_3)) 
      \end{array}
      &  \Gamma
      & \text{type}
      &  \text{order}
      \\
      \hline
      \hline
      M(2,2,K_{\geq 2}) &
      (-1 ; (2,1), (2,1), (K,1) )
      & \text{binary dihedral}
      & D_{K+2}
      & 4 \, K
      \\
      M(2,3,3) &
      (-1 ; (2,1), (3,1), (3,1) )
      & \text{binary tetrahedral}
      & E_6 & 24
      \\
      M(2,3,4) &
      (-1 ; (2,1), (3,1), (4,1)) & \text{binary octahedral} 
      & E_7 &  48
      \\
      M(2,3,5) &
      (-1 ; (2,1), (3,1), (5,1) ) & \text{binary icosahedral} 
      & E_8 & 120
    \end{array}
  \end{equation*}
  \caption{The Seifert manifolds $S^3/\Gamma$ where $\Gamma$ is a
    finite subgroup of SU(2).}
  \label{tab:spherical}
\end{table}

For the  3-manifold $\mathcal{M}$ in Table~\ref{tab:spherical},
the fundamental group becomes
\begin{equation}
  \label{fundamental_solid}
  \pi_1(\mathcal{M})
  \cong \Gamma
  =
  \left\langle
    x_1, x_2, x_3
    ~    \middle| ~
    x_1^{~p_1} =
    x_2^{~p_2} =
    x_3^{~p_3} =
    x_1 \, x_2 \, x_3
    =1
  \right\rangle
\end{equation}
This group is the $(p_1, p_2, p_3)$-triangle group
$T_{p_1, p_2, p_3}$,
and it corresponds to a spherical tessellation due to a
condition~\eqref{condition_spherical}~\cite{JMiln75a}.
It is well known that the group $\Gamma$  is the  symmetry group of
a Platonic solid~\cite{Klein56Book}.
According  to Klein~\cite{Klein56Book},
the $\Gamma$-invariant polynomials on $\mathbb{C}^2$ are  generated by
three fundamental invariants, $x$, $y$, and $z$, and
they  satisfy
$R(x,y,z)=0$
(Table~\ref{tab:surface}),
which basically 
comes  from~\eqref{Brieskorn_surface} after suitable change of variables.
The hypersurface $R(x,y,z)=0$ has a singularity only at
the origin.
This singularity
is the quotient singularity of the hypersurface $\mathbb{C}^2/\Gamma$
in $\mathbb{C}^3$,
and
resolving  these simple singularities gives configuration of rational
curves whose weighted dual graph coincides with the Dynkin diagram of the Lie
algebra as in Table~\ref{tab:surface}
(see, \emph{e.g.}, Ref.~\citen{Slodow83}).

\begin{table}[htbp]
  \centering
  \renewcommand{\arraystretch}{1.4}
  \begin{tabular}{c|c}
    type &
    $R(x, y, z)$
    \\
    \hline \hline
    $D_{K+2}$ &
    $x^2 \, y + y^{K+1} + z^2 = 0$
    \\
    $E_6$ &
    $x^3 + y^4 + z^2=0$
    \\
    $E_7$ &
    $x^3 + x \, y^3 + z^2 =0$
    \\
    $E_8$ &
    $x^3 + y^5 + z^2 = 0$    
  \end{tabular}
  \caption{Type of the Kleinian singularity and hypersurfaces~\cite{Slodow83}.}
  \label{tab:surface}
\end{table}

Our purpose is two-fold.
First we show that the WRT invariant for the spherical Seifert
manifolds $S^3/\Gamma$ can be written in terms of the Eichler integrals
of modular forms with half-integral weight.
This result was first demonstrated by
Lawrence and Zagier in the case of the Poincar{\'e} homology sphere.
Based on this correspondence, we
shall give an exact asymptotic expansion
of the WRT invariant
and study a correspondence with other topological invariants.
In the second part we show that those modular forms are related to the
polyhedral group associated to $\Gamma$, 
and that they construct  a solution of the polyhedral equations.
It suggests that the WRT invariant knows 
the fundamental group in some sense.
This type of correspondence was conjectured in
Ref.~\citen{GuadPilo98a}, and it was checked for a case of lens
space~\cite{SYamad95a}.

This article is constructed as follows.
In Section~\ref{sec:modular}, we present properties of the modular
form.
We define the modular form with weight $3/2$, and give a nearly
modular property of the Eichler integral thereof.
In Section~\ref{sec:WRT} we  give an explicit form  of the WRT invariant
for the Seifert manifolds following Ref.~\citen{LawreRozan99a}.
In Section~\ref{sec:WRT_spherical}, we  show that
the WRT invariant for the spherical  Seifert manifolds
$SU(2)/\Gamma$
in Table~\ref{tab:spherical}
can be written in terms of the Eichler integrals of
the modular form with half-integral weight.
We shall also give an exact asymptotic expansion in $N\to\infty$, and
discuss  the classical topological invariants that appear in this
limit.
In Section~\ref{sec:Platonic}, we study the congruence subgroup, and
we shall reveal that the modular form
is related to the polyhedral group.
The last section is devoted to concluding remarks and discussions.

\section{Preliminaries}
\label{sec:modular}

Throughout this article, we set
\begin{equation*}
  q=\exp \left( 2 \, \pi \, \I \, \tau \right)
\end{equation*}
where $\tau$ is in the upper half plane,
$\tau\in \mathbb{H}$.
We use the Dedekind $\eta$-function,
\begin{equation}
  \label{Dedekind_eta}
  \eta(\tau)
  =
  q^{\frac{1}{24}} \,
  \prod_{n=1}^\infty
  \left( 1-q^n \right)
\end{equation}
which is a modular form with weight $1/2$
satisfying
\begin{equation}
  \begin{gathered}
    \eta(-1 / \tau)
    =
    \sqrt{
      \frac{\tau}{\I}
    } \, \eta(\tau)
    \\[2mm]
    \eta(\tau+1)
    =
    \E^{\frac{1}{12} \pi \I} \,
    \eta(\tau)
  \end{gathered}
\end{equation}

Another important family of the modular forms is
the 
(normalized) Eisenstein  series (see, \emph{e.g.},
Ref.~\citen{Koblitz93Book})
\begin{equation}
  \label{Eisenstein_series}
  E_k(\tau)
  =
  \frac{1}{2 \, \zeta(k)} \,
  \sum_{\substack{
      (m,n) \in \mathbb{Z}^2
      \\
      (m,n) \neq (0,0)
    }}
  \frac{1}{
    \left( m \, \tau + n \right)^k
  }
\end{equation}
Here $k$ is even integer greater than $2$,
and the Riemann $\zeta$-function is
\begin{equation*}
  \zeta(k) = \sum_{n=1}^\infty \frac{1}{n^k}
\end{equation*}
Note that the $\zeta$-function for even $k$ is given by
\begin{equation*}
  \zeta(k)
  =
  - \frac{
    \left( 2 \, \pi \, \I \right)^k
  }{2 \, k!} \, B_k
  \quad
  \text{
    for even $k_{\geq 2}$
  }
\end{equation*}
where
$B_k$ is the $k$-th Bernoulli number,
\begin{equation*}
  \frac{t}{\E^t - 1}
  =
  \sum_{k=0}^\infty
  B_k \,
  \frac{t^k}{k!}
\end{equation*}
The Eisenstein series is a modular form with weight $k$,
\begin{equation}
  E_k(-1/\tau)
  =
  \tau^k \, E_k(\tau)
\end{equation}
and
it   has a Fourier expansion
\begin{equation}
  E_k(\tau)
  =
  1 - \frac{2 \, k}{B_k}
  \sum_{n=1}^\infty
  \sigma_{k-1}(n) \, q^n
\end{equation}
Here the arithmetic function $\sigma_k(n)$ is  a sum of the $k$-th
powers of the positive divisors of $n$,
\begin{equation*}
  \sigma_k(n)
  =
  \sum_{d | n} d^k
\end{equation*}
and  the Fourier expansion can be rewritten
in the form of the
Lambert series 
\begin{equation*}
  E_k(\tau)
  =
  1- \frac{2 \, k}{B_k} \sum_{n=1}^\infty
  \frac{n^{k-1} \, q^n}{1-q^n}
\end{equation*}
The cusp form with weight $12$ is the Ramanujan $\Delta$-function
\begin{equation}
  \Delta(\tau)=
  \left(
    \eta(\tau)
  \right)^{24}
\end{equation}
and it
is given from  the Eisenstein series as
\begin{equation}
  \label{E_E_Delta}
  \left( E_4(\tau) \right)^3
  -
  \left( E_6(\tau) \right)^2
  =
  1728 \,
  \Delta(\tau)
\end{equation}

Besides  the Dedekind $\eta$-function~\eqref{Dedekind_eta}, we
make use of another family of the modular form
with half-integral weight~\cite{KHikami03a}.
For $P\in \mathbb{Z}_{>0}$ and $a\in\mathbb{Z}$ we set
\begin{equation}
  \Psi_{ P}^{(a)}(\tau)
  =
  \frac{1}{2}
  \sum_{n \in \mathbb{Z}} n \,
  \psi_{2 P}^{(a)}(n) \,
  q^{\frac{n^2}{4 P}}
\end{equation}
where $\psi_{2 P}^{(a)}(n)$ is an odd periodic function with modulus
$2 \, P$;
\begin{equation}
  \psi_{2 P}^{(a)}(n)
  =
  \begin{cases}
    \pm 1 & \text{for $n \equiv \pm a \mod 2  \, P$}
    \\[2mm]
    0 & \text{otherwise}
  \end{cases}
\end{equation}
These $q$-series are related to the characters of the affine Lie
algebra $\widehat{su(2)}$
(see,  \emph{e.g.}, Refs.~\citen{Kac90,DiFrMathSene97}).
We see that this family of  $q$-series
is a vector modular form with weight $3/2$;
under the modular $S$- and $T$-transformations,
\begin{align*}
  S & :
  \tau \to - \frac{1}{\tau}
  \\[2mm]
  T & :
  \tau \to \tau+1
\end{align*}
satisfying
\begin{equation*}
  S^2 = \left( S \, T \right)^3 = 1
\end{equation*}
it  transforms  as
\begin{equation}
  \label{Psi_modular}
\begin{gathered}
  \Psi_{P}^{(a)}(\tau)
  =
  \left(
    \frac{\I}{\tau}
  \right)^{\frac{3}{2}} \,
  \sum_{b=1}^{P-1}
  \mathbf{M}(P)^a_b \,
  \Psi_P^{(b)}(-1/\tau)
  \\[2mm]
  \Psi_P^{(a)}(\tau + 1)
  =
  \E^{\frac{a^2}{2 P} \pi \I}
  \,
  \Psi_{P}^{(a)}(\tau)
\end{gathered}
\end{equation}
Here  $\mathbf{M}(P)$ is a $(P-1)\times (P-1)$  matrix whose elements are
\begin{equation}
  \mathbf{M}(P)^a_b =
  \sqrt{\frac{2}{P}} \,
  \sin\left( \frac{a\, b}{P} \, \pi \right)
\end{equation}

Following  Ref.~\citen{LawrZagi99a},
we define
the Eichler integral of this family of the modular forms with
half-integral weight  by
(see also Ref.~\citen{KHikami03a})
\begin{equation}
  \label{define_Eichler}
  \widetilde{\Psi}_P^{(a)}(\tau)
  =
  \sum_{n=0}^\infty
  \psi_{2 P}^{(a)}(n) \,
  q^{\frac{n^2}{4 P}}
\end{equation}
This can be regarded as a \emph{half-integration} of
$\Psi_P^{(a)}(\tau)$ with respect to $\tau$.
A limiting values of the Eichler integral in
$\tau \to \frac{M}{N} \in \mathbb{Q}$ can be computed by use of the Mellin
transformation, and
we have~\cite{KHikami03a}
\begin{gather}
  \label{Eichler_over_N}
  \widetilde{\Psi}_P^{(a)}(1/N)
  =
  -\sum_{k=0}^{2 P N}
  \psi_{2 P}^{(a)}(k) \,
  \E^{\frac{k^2}{2 P N} \pi \I} \,
  B_1
  \left(
    \frac{k}{2 \, P \, N}
  \right)
  \\[2mm]
  \label{Eichler_at_N}
  \widetilde{\Psi}_P^{(a)}(N)
  =
  \left(
    1- \frac{a}{P}
  \right) \,
  \E^{\frac{a^2}{2 P} \pi \I N}
\end{gather}
where
$N\in \mathbb{Z}$, and
$B_k(x)$ denotes the $k$-th Bernoulli polynomial defined by
\begin{equation*}
  \frac{
    t \, \E^{x \, t}
  }{
    \E^t - 1
  }
  =
  \sum_{k=0}^\infty
  \frac{B_k(x)}{k!} \, t^k
\end{equation*}

{}From the topological viewpoint,
a  limiting value~\eqref{Eichler_over_N} in $\tau \to 1/N$
is related to the specific value of the $N$-colored
Jones polynomial for torus links $\mathcal{T}_{2,2 P}$
with  $P>0$~\cite{KHikami03a}.
Explicitly we have
\begin{equation}
  \label{Kashaev_link}
  \left\langle \mathcal{T}_{2, 2 P} \right\rangle_N
  =
  P \, N \,
  \E^{- \frac{(P-1)^2}{2 P N} \pi \I} \,
  \widetilde{\Psi}_{P}^{(P-1)}(1/N)
\end{equation}
Here $\langle \mathcal{K} \rangle_N$ is  Kashaev's invariant for a
knot $\mathcal{K}$~\cite{Kasha95}, and it coincides with a specific
value of the $N$-colored Jones polynomial
$J_N(q; \mathcal{K})$ as~\cite{MuraMura99a}
\begin{equation}
  \left\langle \mathcal{K} \right\rangle_N
  =
  J_N
  \left(
    \frac{2 \, \pi \, \I}{N} ; \mathcal{K}
  \right)
\end{equation}
where we have normalized the quantum invariant s.t.
\begin{equation*}
  \left\langle \text{unknot} \right\rangle_N = 1
\end{equation*}
Topological meaning of
other  Eichler integrals
$\widetilde{\Psi}_{P}^{(a)}(1/N)$ with $a \neq P-1$
is not clear,
and we show hereafter that some of them are related to the WRT
invariant for the spherical Seifert manifolds $M(p_1,p_2, p_3)$.

A crucial property  of the Eichler integral~\eqref{define_Eichler} is
that it is nearly modular~\cite{LawrZagi99a,DZagie01a}.
For $N\in \mathbb{Z}_{>0}$,
we have an exact asymptotic expansion in $N\to\infty$ as
\begin{equation}
  \label{Psi_nearly_modular}
  \widetilde{\Psi}_P^{(a)}(1/N)
  +
  \sqrt{\frac{N}{\I}} \,
  \sum_{b=1}^{P-1}
  \mathbf{M}(P)^a_b \,
  \widetilde{\Psi}_P^{(b)}(-N)
  \simeq
  \sum_{k=0}^\infty
  \frac{
    L\left(
      -2 \, k , \psi_{2 P}^{(a)}
    \right)
  }{k!} \,
  \left(
    \frac{\pi \, \I}{2 \, P \, N}
  \right)^k
\end{equation}
where
$\widetilde{\Psi}_P^{(a)}(1/N)$ and
$\widetilde{\Psi}_P^{(b)}(N)$ are given in~\eqref{Eichler_over_N}
and~\eqref{Eichler_at_N}.
The Dirichlet $L$-function $L\left(s,\psi_{2P}^{(a)}\right)$  at
negative integers 
$s=-k 
$ is given by
\begin{equation}
  L\left( -k, \psi_{2 P}^{(a)}\right)
  =
  -
  \frac{\left( 2 \, P \right)^k}{k+1} \,
  \sum_{n=1}^{2 P}
  \psi_{2 P}^{(a)}(n) \, B_{k+1} \left( \frac{n}{2 \, P} \right)
\end{equation}
See Refs.~\citen{LawrZagi99a,DZagie01a,KHikami03a} for proof.
We should remark that
the generating function of the $L$-functions at negative integers,
$L\left( -2 \, k , \psi_{2 P}^{(a)} \right)$ for $0<a<P$,
is
given by
\begin{equation}
  \frac{\sh( (P-a) \, z )}{\sh(P \, z)}
  =
  \sum_{k=0}^\infty
  \frac{
    L
    \left(
      -2 \, k , \psi_{2 P}^{(a)}
    \right)
  }{
    (2 \, k) !
  } \,
  z^{2 k}
\end{equation}

To close this section,
we recall the
Gauss sum reciprocity formula~\cite{LCJeff92a,Chandrasek85},
\begin{equation}
  \label{Gauss_reciprocity}
  \sum_{n \mod N}
  \mathrm{e}^{\frac{\pi \mathrm{i}}{N} M n^2 + 2 \pi \mathrm{i} k n}
  =
  \sqrt{
    \left|
      \frac{N}{M}
    \right|
  } \,
  \mathrm{e}^{\frac{\pi \mathrm{i}}{4} \sign(N M)}
  \sum_{n \mod M}
  \mathrm{e}^{
    -\frac{\pi \mathrm{i}}{M}
    N (n+k)^2
  }
\end{equation}
where $N, M\in \mathbb{Z}$
with
$N \, k \in \mathbb{Z}$ and
$N \, M$ being even.
This can be derived based on the transformation law of the theta
series.

\section{Witten--Reshetikhin--Turaev Invariant}
\label{sec:WRT}

The explicit form of the WRT invariant for the Seifert manifolds is 
given  by the  method of Reshetikhin and Turaev~\cite{ResheTurae91a}.
Based on a surgery description of  3-manifold $\mathcal{M}$, we
can compute the SU(2) WRT invariant using the colored Jones polynomial
for link.
The SU(2) WRT invariant for the
Seifert manifolds has been extensively studied
(see, \emph{e.g.}, Refs.~\citen{RLawre95a,RLawre96a,LawreRozan99a,Rozan95a,LRozansky95a,Rozan96c,Rozan96e}),
and we note the known   result as follows;

\begin{prop}[\cite{LawreRozan99a}]
Let $\mathcal{M}$ be the Seifert manifold
${M}\left(
  0; 
  (p_1, q_1), (p_2, q_2), (p_3, q_3)
\right)$.
Then we have
\begin{multline}
  \label{Lawrence_Rozansky_form}
  \E^{\frac{2 \pi \I}{N} \left(
      \frac{\phi}{4} - \frac{1}{2}
    \right)}
  \, \left(
    \E^{\frac{2 \pi \I}{N}} - 1
  \right) \cdot
  \tau_N(\mathcal{M})
  \\
  =
  \frac{\E^{\frac{\pi \I}{4}}}{\sqrt{2 \, N \, p_1 \, p_2 \, p_3}}
  \sum_{k_0=1}^{N-1}
  \sum_{n_j \mod p_j}
  \frac{1}{
    \E^{\frac{\pi \I}{N} k_0} -
    \E^{- \frac{\pi \I}{N} k_0}
  } \,
  \\
  \times
  \prod_{j=1}^3
  \E^{- \frac{\pi \I}{2 N} \frac{q_j}{p_j}
    \left( k_0 + 2  N n_j \right)^2
  } \,
  \left(
    \E^{\frac{\pi \I}{N p_j}
      \left( k_0 + 2 N n_j \right)
    }
    -
    \E^{- \frac{\pi \I}{N p_j}
      \left( k_0 + 2 N n_j \right)
    }
  \right)
\end{multline}
Here we have set
\begin{equation}
  \phi
  =
  \sum_{j=1}^3
  \left(
    12 \, s(q_j, p_j) -
    \frac{q_j}{p_j}
  \right) 
  +3
\end{equation}
where
$s(b,a)$ is the Dedekind sum (see, \emph{e.g.},
Ref.~\citen{RademGross72})
\begin{equation}
  s(b,a)
  =
  \sign(a)
  \sum_{k=1}^{
    |a| -1
  }
  \Bigl( \! \Bigl( \frac{k}{a} \Bigr) \! \Bigr) \cdot
  \Bigl( \! \Bigl( \frac{k \, b}{a} \Bigr) \! \Bigr) 
\end{equation}
with
\begin{equation*}
  ( \! (x) \! )
  =
  \begin{cases}
    \displaystyle
    x - \left\lfloor x \right\rfloor - \frac{1}{2}
    &
    \text{if $x \not\in \mathbb{Z}$}
    \\[2mm]
    0 &
    \text{if $x \in \mathbb{Z}$}
  \end{cases}
\end{equation*}
and
$\left\lfloor x \right\rfloor$ is the greatest integer not exceeding $x$.
\end{prop}

It is known that
the SU(2) WRT invariant for a 3-manifold $\mathcal{M}$
can be  factorized~\cite{KirbMelv91a} as
\begin{equation}
  \label{factorize_tau}
  \tau_N(\mathcal{M})
  =
  \begin{cases}
    \tau_3(\mathcal{M}) \, 
    \tau_N^{SO(3)}(\mathcal{M})
    &
    \text{for $N =3 \mod 4$}
    \\[2mm]
    \overline{
      \tau_3(\mathcal{M})
    } \, 
    \tau_N^{SO(3)}(\mathcal{M})
    &
    \text{for $N =1 \mod 4$}
  \end{cases}
\end{equation}
where $\tau_N^{SO(3)}(\mathcal{M})$ is the SO(3) WRT
invariant, and
\begin{equation}
  \label{define_tau3}
  \tau_3(\mathcal{M})
  =
  (1 + \I)^{\sigma_+} \,
  (1 - \I)^{\sigma_-} \,
  \sum_{
    \vv{x} \in
    \left(
      \mathbb{Z}/2 \mathbb{Z}
    \right)^\ell
  }
  \I^{ {}^t \vv{x} \mathbf{L} \vv{x}}
\end{equation}
where $\mathbf{L}$ is the linking matrix of a link which presents a
surgery description of $\mathcal{M}$,
and $\sigma_\pm$ denote the number of positive/negative
eigenvalues of the linking matrix $\mathbf{L}$.
Studied in detail~\cite{KirbMelv91a} is the condition for the manifold
$\mathcal{M}$ that 
$\tau_3(\mathcal{M})=0$.

It is remarked that the Dedekind sum can also be written
as~\cite{RademGross72}
\begin{equation*}
  s(b,a)
  =
  \frac{1}{4 \, |a|}
  \sum_{k=1}^{
    |a| -1 }
  \cot
  \left(\frac{k}{a} \, \pi \right) \,
  \cot
  \left(
    \frac{k \, b}{a} \, \pi \right)
\end{equation*}
and that it satisfies
\begin{gather*}
  s(-b,a) = - s(b,a)
  \\[2mm]
  s(b,a) =
  s(b^\prime , a)
  \quad
  \text{for $b \, b^\prime \equiv 1 \pmod a$}
\end{gather*}
It is well known that the Dedekind sum is related to the
Casson--Walker invariant, which naively denotes  the number of the
irreducible SU(2) representation of the fundamental group
$\pi_1(\mathcal{M})$.
Explicitly the Casson--Walker invariant $\lambda_{CW}(\mathcal{M})$
for
$\mathcal{M}=
M
\left(
b;(a_1,b_1), (a_2, b_2), (a_3, b_3)
\right)$
is given by~\cite{FukuMatsSaka90,FintuStern90a}
\begin{multline}
  \lambda_{CW}(\mathcal{M})
  =
  \frac{a_1 \, a_2 \, a_3}{8} \,
  \Biggl(
    \frac{\sign(e(\mathcal{M}))}{3} \,
    \left(
      -1 + \sum_{j=1}^3 \frac{1}{a_j^{~2}}
    \right)
    \\
    +
    \frac{e(\mathcal{M}) \,
      \left|
        e(\mathcal{M})
      \right|
    }{3}
    - e(\mathcal{M})
    - 4 \, \left| e(\mathcal{M}) \right| \,
    \sum_{j=1}^3
    s(b_j,a_j)
  \Biggr)
\end{multline}
where $e(\mathcal{M})$ is the Euler characteristic
\begin{equation*}
  e(\mathcal{M})
  =
  b+\sum_{j=1}^3 \frac{b_j}{a_j}
\end{equation*}

Interest in asymptotic behavior of the WRT invariant is motivated
by Witten's original results that the saddle point of the Chern--Simons
path integral
in the large $N$ limit
is given by contribution coming solely from the flat connections,
and that (classical)
topological invariants should appear in this  limit.
The asymptotic behavior of the WRT invariant
for $3$-manifold $\mathcal{M}$ is expected to
be~\cite{FreeGomp91a,Rozan95a}
\begin{equation}
  \label{Z_asymptotics}
  Z_k(\mathcal{M})
  \sim
  \frac{1}{2} \,
  \E^{- \frac{3}{4} \pi \I}
  \sum_\alpha
  \sqrt{T_\alpha(\mathcal{M})} \,
  \E^{- \frac{2 \pi \I}{4} I_\alpha} \,
  \E^{2 \pi \I  (k+2) \CS(A_\alpha) }
\end{equation}
Here $Z_k(\mathcal{M})$ is the partition function due to Witten's
normalization,
\begin{equation}
  \label{Witten_partition_function}
  Z_k(\mathcal{M})
  =
  \frac{
    \tau_{k+2}(\mathcal{M})
  }{
    \tau_{k+2} \left(S^2 \times S^1 \right)
  }
\end{equation}
where 
\begin{equation*}
  \tau_N \left( S^2 \times S^1 \right)
  =
  \sqrt{
    \frac{N}{2}
  } \,
  \frac{1}{
    \sin \left( \pi / N \right)
  }
\end{equation*}
The index  $\alpha$ ranges over all gauge equivalence classes of flat
connections.
The Reidemeister torsion, spectral flow, and the
Chern--Simons invariant are respectively denoted by
$T_\alpha$, $I_\alpha$, and $\CS(A)$.

In the case of the Seifert manifold
$M(p_1, p_2, p_3)$,
the explicit values of the torsion and the Chern--Simons invariant are
known.
The Reidemeister torsion is given
by~\cite{DFreed92a}
\begin{equation}
  \label{torsion_ell}
  \sqrt{T_\alpha}
  =
  \prod_{j=1}^3
  \frac{2}{\sqrt{p_j}}
  \left|
    \sin \left( \frac{q_j^\prime \,\ell_j}{p_j} \, \pi \right)
  \right|
\end{equation}
where
$q_j \, q_j^\prime \equiv 1 \mod p_j$.
An integer $\ell_j$ satisfying
$0<\ell_j<p_j$ parametrizes the
irreducible SU(2)
representation $\rho$ of the fundamental
group~\eqref{fundamental_solid}, and we have
\begin{equation*}
  \rho(x_j) \sim
  \begin{pmatrix}
    \E^{\frac{\ell_j}{p_j} \pi \I} &
    \\
    & \E^{- \frac{\ell_j}{p_j} \pi \I}
  \end{pmatrix}
\end{equation*}
up to conjugation.
Corresponding to this representation $\rho$, the Chern--Simons
invariant is given by~\cite{Auckl94a,FintuStern90a,KirkKlas90a,HNishi98}
\begin{equation}
  \label{CS_ell}
  \CS(A_\alpha)
  =
  - \frac{1}{4} 
  \sum_{j=1}^3
  \frac{q_j}{p_j} \,
  \ell_j^{~2}
  \mod 1
\end{equation}


\section{WRT Invariant for the Spherical Seifert Manifolds}
\label{sec:WRT_spherical}

We shall clarify the relationship between the WRT invariant for the
spherical Seifert manifolds $S^3/\Gamma$ 
in Table~\ref{tab:spherical} and the Eichler integral
$\widetilde{\Psi}_P^{(a)}(1/N)$
of  
modular forms
with weight $3/2$.
The
expression~\eqref{Lawrence_Rozansky_form} can be simplified into that
with a unique  sum for a case of the homology sphere~\cite{LawreRozan99a},
but in our case it is necessary to treat each case one by one.



\subsection{\mathversion{bold}
  $M(2,3,3)$
}

Let $\mathcal{E}_6$ be ${M}(2,3,3)$.
The Euler characteristic is given by
\begin{gather*}
  e(\mathcal{E}_6) = \frac{1}{6}
\end{gather*}
The surgery description given in~\eqref{Seifert_surgery} can be
transformed into the following form by  the Kirby move;
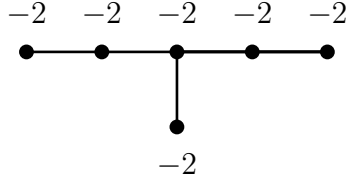
\begin{equation*}
  \raisebox{-1.6cm}{
    \begin{pspicture}(-2,-2)(4,1)
      \psdots[dotsize=4pt 2](-1,0)(0,0)(1,0)(2,0)(3,0)(1,-1)
      \pscustom[linewidth=1.0pt]{
        \psline{-}(-1,0)(0,0)(1,0)(2,0)(3,0)
        \psline{-}(1,0)(1,-1)
      }
      \rput{0}(-1,0.5){$-2$}
      \rput{0}(0,0.5){$-2$}
      \rput{0}(1,0.5){$-2$}
      \rput{0}(2,0.5){$-2$}
      \rput{0}(3,0.5){$-2$}
      \rput{0}(1,-1.5){$-2$}
    \end{pspicture}
  }
\end{equation*}
This is nothing but the Dynkin diagram for the Lie algebra $E_6$.

\begin{prop}
  \label{prop:E6}
  The SU(2) WRT invariant for $\mathcal{E}_6$ is written as a sum of
  the Eichler integrals $\widetilde{\Psi}_6^{(a)}(1/N)$ as
  \begin{multline}
    \label{WRT_E6}
    \E^{\frac{13}{12 N} \pi \I} \,
    \left(
      \E^{\frac{2 \pi\I}{N}} - 1
    \right) \cdot
    \tau_N\left(
      \mathcal{E}_6
    \right)
    \\
    =
    \frac{1 + 2 \,  \E^{\frac{2}{3} \pi \I N}}{\sqrt{3}} \,
    \E^{\frac{\pi  \I}{12 N}}
    -
    \frac{1 + 2 \, \E^{\frac{2}{3} \pi \I N}}{2 \sqrt{3}} \,
    \left(
      \widetilde{\Psi}_6^{(1)}(1/N)
      +
      \widetilde{\Psi}_6^{(5)}(1/N)
    \right)
    -
    \frac{1 - \E^{\frac{2}{3} \pi \I N}}{ \sqrt{3}} \,
    \widetilde{\Psi}_6^{(3)}(1/N)
  \end{multline}
\end{prop}

\begin{proof}
In the case of the manifold $\mathcal{E}_6$
we have $\vv{p}=(2,3,3)$ and $\vv{q}=(-1,1,1)$
in~\eqref{Lawrence_Rozansky_form}.
To rewrite this expression
in terms of the Eichler integrals,
we note
that the summand of the right hand side
of~\eqref{Lawrence_Rozansky_form} is invariant~\cite{LawreRozan99a}
under
\begin{itemize}
\item $k_0 \to k_0 + 2 \, N$ and
  $\forall n_j \to n_j -1$
\item
  $n_j \to n_j + p_j$
\end{itemize}
With above symmetries,
the
sum
$\sum_{k_0=1}^{N-1} \sum_{n_j \mod p_j}$ can be replaced with a sum
$\sum_{\substack{
    k_0 = a + 2 N n
    \\
    1 \leq a \leq N-1
    \\
    0 \leq n \leq 5
  }}
\sum_{n_3 = 0}^2$ and   $n_1=n_2=0$.
After taking a sum over $n_3$ explicitly, we get
\begin{multline*}
  \text{l.h.s. of \eqref{WRT_E6}}
  =
  \frac{\E^{\frac{\pi \I}{4}}}{6  \, \sqrt{N}}
  \sum_{\substack{
      k_0=a + 2 N n
      \\
      1 \leq a \leq N-1
      \\
      0 \leq n \leq 5}}
  \frac{
    \left(
      \E^{\frac{\pi \I}{2 N} k_0} -
      \E^{- \frac{\pi \I}{2 N} k_0}
    \right) \,
    \left(
      \E^{\frac{\pi \I}{3 N} k_0} -
      \E^{- \frac{\pi \I}{3 N} k_0}
    \right)
  }{
    \E^{\frac{\pi \I}{ N} k_0} -
    \E^{- \frac{\pi \I}{ N} k_0}
  }
  \,
  \E^{\frac{- \pi \I}{12 N} k_0^{~2}}
  \\
  \times
  \left(
    \E^{\frac{\pi \I}{3 N} k_0} \,
    \left(
      1 + \E^{\frac{2}{3} \pi \I ( 1 - N - k_0)}
      + \E^{\frac{2}{3} \pi \I ( -1 - N + k_0)}
    \right)
  \right.
  \\
  \left.
    -
    \E^{-\frac{\pi \I}{3 N} k_0} \,
    \left(
      1 + \E^{\frac{2}{3} \pi \I ( 1 - N + k_0)}
      + \E^{\frac{2}{3} \pi \I ( -1 - N - k_0)}
    \right)
  \right)
\end{multline*}
As the summand of the above expression is invariant under
$k_0
\rightarrow 12 \, N - k_0$,
we obtain
\begin{multline*}
  \text{l.h.s. of \eqref{WRT_E6}}
  =
  \frac{\E^{\frac{\pi \I}{4}}}{12 \, \sqrt{N}}
  \sum_{\substack{
      k_0=0
      \\
      N \nmid k_0}}^{12 \, N -1}
  \frac{
    \left(
      \E^{\frac{\pi \I}{2 N} k_0}
      -
      \E^{- \frac{\pi \I}{2 N} k_0}
    \right) \,
    \left(
      \E^{\frac{\pi \I}{3 N} k_0}
      -
      \E^{- \frac{\pi \I}{3 N} k_0}
    \right) 
  }{
    \E^{\frac{\pi \I}{N} k_0}
    -
    \E^{- \frac{\pi \I}{N} k_0}
  } \,
  \E^{-\frac{\pi \I}{12 N} k_0^{~2}} 
  \\
  \times
  \left(
    \E^{\frac{\pi \I}{3 N} k_0} \,
    \left(
      1 + \E^{-\frac{2}{3} \pi \I N} \,
      \left(
        3 \, \delta_{3 \mid ( k_0 -1 )} - 1
      \right)
    \right)
    -
    \E^{-\frac{\pi \I}{3 N} k_0} \,
    \left(
      1 + \E^{-\frac{2}{3} \pi \I N} \,
      \left(
        3 \, \delta_{3 \mid ( k_0 +1)} - 1
      \right)
    \right)
  \right)
\end{multline*}
where we have used
\begin{equation*}
  1 + \E^{\frac{2 \pi \I}{3} n }
  + \E^{- \frac{2 \pi \I}{3} n }
  =
  3 \, \delta_{3 \mid n}
\end{equation*}
We then introduce the even periodic function $\chi_{12}(n)$ with
modulus $12$ by
\begin{equation*}
  \begin{array}{c|ccccc}
    n \mod 12 & 1 & 5 & 7 &11 &  \text{others}
    \\
    \hline
    \chi_{12}(n) & 1 & -1 & -1 & 1 & 0
  \end{array}
\end{equation*}
whose generating function is
\begin{equation}
  \label{define_chi12}
  \frac{
    \left( t^2 - t^{-2} \right) \,
    \left( t^3 - t^{-3} \right) 
  }{
    t^6 - t^{-6}
  }
  = -
  \sum_{n=0}^\infty
  \chi_{12}(n) \, t^n
\end{equation}
Using this, we have
\begin{multline*}
  \text{l.h.s. of \eqref{WRT_E6}}
  =
  - \frac{\E^{\frac{\pi \I}{4}}}{12 \, \sqrt{N}}
  \lim_{t\searrow 0} \sum_{n=0}^\infty
  \chi_{12}(n) \,
  \E^{-n t} \,
  \sum_{      k_0=0}^{12 \, N -1}
  \E^{\frac{\pi \I}{6 N} k_0 n - \frac{\pi \I}{12 N} k_0^{~2}}
  \\
  \times
  \left(
    \E^{\frac{\pi \I}{3 N} k_0} \,
    \left(
      1 + \E^{-\frac{2}{3} \pi \I N} \,
      \left(
        3 \, \delta_{3 \mid ( k_0 -1 )} - 1
      \right)
    \right)
    -
    \E^{-\frac{\pi \I}{3 N} k_0} \,
    \left(
      1 + \E^{-\frac{2}{3} \pi \I N} \,
      \left(
        3 \, \delta_{3 \mid ( k_0 +1)} - 1
      \right)
    \right)
  \right)
\end{multline*}
where we have used a fact that the sum for $N \mid k_0$ is zero.
We apply the Gauss sum reciprocity formula~\eqref{Gauss_reciprocity},
and then use an identity
\begin{equation*}
  \sum_{k \mod 3}
  \E^{\frac{4}{3} N \pi \I
    \left(
      k + \frac{x}{4 N}
    \right)^2
  }
  =
  \E^{\frac{x^2}{12 N} \pi \I} \,
  \left(
    1+ \E^{-\frac{2}{3} N \pi \I}
    \left(
      3 \delta_{3 \mid x} - 1
    \right)
  \right)
\end{equation*}
for $x \in \mathbb{Z}$.
After some computations, we find
\begin{multline*}
  \text{l.h.s. of \eqref{WRT_E6}}
  \\
  =
  \frac{-1}{\sqrt{12}} 
  \lim_{t \searrow 0}
  \sum_{n=0}^\infty \E^{-n t} \, \chi_{12}(n) \,
  \Biggl(
    \E^{\frac{(n+2)^2}{12 N} \pi \I} \,
    \left(
      \left(
        1+2 \, \E^{-\frac{4}{3} \pi \I N}
      \right) \, \delta_{3 \mid ( n+1)}
      +
      \left(
        1- \E^{-\frac{4}{3} \pi \I N}
      \right) \, \delta_{3 \mid ( n+2)}
    \right)
    \\
    -
    \E^{\frac{(n-2)^2}{12 N} \pi \I} \,
    \left(
      \left(
        1- \E^{-\frac{4}{3} \pi \I N}
      \right) \, \delta_{3 \mid ( n+1)}
      +
      \left(
        1+2  \, \E^{-\frac{4}{3} \pi \I N}
      \right) \, \delta_{3 \mid ( n+2)}
    \right)
  \Biggr)
  \\
  =
  \frac{1}{\sqrt{12}}
  \sum_{k=0}^{12 N -1}
  B_1 \left(\frac{k}{12 \, N}\right) \,
  \Biggl( 
    \E^{ \frac{(k+2)^2}{12 N} \pi \I} \,
    \chi_{12}^{(-)} (k) \,
    \left(1+2 \, \E^{- \frac{4}{3} \pi \I N} \right)
    +
    \E^{ \frac{(k+2)^2}{12 N} \pi \I} \,
    \chi_{12}^{(+)}  (k)\,
    \left(1- \E^{- \frac{4}{3} \pi \I N} \right)
    \\
    -
    \E^{ \frac{(k-2)^2}{12 N} \pi \I} \,
    \chi_{12}^{(-)}  (k)\,
    \left(1- \E^{- \frac{4}{3} \pi \I N} \right)
    -
    \E^{ \frac{(k-2)^2}{12 N} \pi \I} \,
    \chi_{12}^{(+)}  (k)\,
    \left(1+2 \, \E^{- \frac{4}{3} \pi \I N} \right)
  \Biggr)
\end{multline*}
Here  we have defined
the periodic functions $\chi_{12}^{(\pm)}(k)$
by
\begin{gather*}
  \begin{array}{c|ccc}
    n \mod 12 & 1  & 7 &  \text{others}
    \\
    \hline
    \chi_{12}^{(+)}(n) & 1 & -1  & 0
  \end{array}
  \\[2mm]
  \begin{array}{c|ccc}
    n \mod 12 & 5  & 11 &  \text{others}
    \\
    \hline
    \chi_{12}^{(-)}(n) & -1 & 1  & 0
  \end{array}
\end{gather*}
which  satisfy
$\chi_{12}(k)=
\chi_{12}^{(+)}(k)+\chi_{12}^{(-)}(k)$.
Finally we  shift  a sum $k\to k \pm 2$, and
use  a relationship between the periodic functions
$\chi_{12}^{(\pm)}(n)$ and $\psi_{12}^{(a)}(n)$, such as
$\chi_{12}^{(\pm)}(n \mp 2)=\pm \psi_{12}^{(3)}(n)$,
$\chi_{12}^{(+)}(n)=\chi_{12}^{(-)}(n-2)$,
and
$\chi_{12}^{(+)}(n) - \chi_{12}^{(-)}(n) =
\psi_{12}^{(1)}(n)+\psi_{12}^{(5)}(n)
$.
Reforming a sum and using an expression~\eqref{Eichler_over_N},
we get the assertion of the
proposition.
\end{proof}

\begin{coro}
  Exact asymptotic expansion  of the WRT invariant in $N\to\infty$ is
  given by
  \begin{multline}
    \label{exact_asymptotic_E6}
    \E^{\frac{13}{12 N} \pi \I} \,
    \left(
      \E^{\frac{2 \pi\I}{N}} - 1
    \right) \cdot
    \tau_N\left( \mathcal{E}_6
    \right)
    \simeq
    \sqrt{\frac{N}{\I}} \,
    \E^{- \frac{1}{12} \pi \I N}
    +
    \frac{
      1+ 2 \, \E^{\frac{2}{3} \pi \I N}
    }{\sqrt{3}}
    \,
    \E^{\frac{\pi \I}{12 N}}
    \\
    -
    \sum_{k=0}^\infty
    \frac{
      1
    }{
      k!
    } \,
    \left(
      \frac{
        1+ 2 \, \E^{\frac{2}{3} \pi \I N}
      }{2 \sqrt{3}} \,
      L\left(
        -2 \, k , \psi_{12}^{(1)}+ \psi_{12}^{(5)}
      \right)
      +
      \frac{
        1 -  \E^{\frac{2}{3} \pi \I N}
      }{ \sqrt{3}} \,
      L      \left(
        -2 \, k , \psi_{12}^{(3)}
      \right)
    \right) \,
    \left(
      \frac{\pi \,  \I}{12 \, N}
    \right)^k
  \end{multline}
  where  the generating functions of the $L$-functions are given by
  \begin{gather*}
    \frac{
      \ch (2 \, z)
    }{
      \ch(3 \, z)
    }
    =
    \sum_{k=0}^\infty
    \frac{
      L \left(
        - 2 \, k ,  \psi_{12}^{(1)} + \psi_{12}^{(5)}
      \right)
    }{
      (2 \, k) !
    } \,
    z^{2 k}
    \\[2mm]
    \frac{
      1
    }{
      2 \, \ch(3 \, z)
    }
    =
    \sum_{k=0}^\infty
    \frac{
      L \left(
        - 2 \, k ,  \psi_{12}^{(3)}
      \right)
    }{
      (2 \, k) !
    } \,
    z^{2 k}
  \end{gather*}
  We thus have a  dominating term of the Witten partition function in
  $N\to\infty$ as
  \begin{equation}
    \label{E6_asymptotics}
    Z_{N-2}(\mathcal{E}_6)
    \sim
    \frac{1}{2} \, \E^{-\frac{3}{4} \pi \I} \cdot \sqrt{2} \,
    \E^{-\frac{1}{12} \pi \I N}
  \end{equation}
\end{coro}

\begin{proof}
  We apply~\eqref{Psi_nearly_modular},
  and
  we obtain~\eqref{exact_asymptotic_E6} immediately.
  As a result,
  the dominating terms of
  the partition function $Z_k(\mathcal{E}_6)$, which is defined
  in~\eqref{Witten_partition_function},
  can be given.  
\end{proof}

The torsion and the Chern--Simons invariant are respectively computed
from~\eqref{torsion_ell} and~\eqref{CS_ell} by setting
surgery data,
$\vv{p}=(2,3,3)$ and $\vv{q}=(-1,1,1)$,
for $\mathcal{E}_6$.
By choosing $\vv{\ell}=(1,1,1)$, we have
\begin{equation}
  \begin{gathered}
    \sqrt{T_\alpha} = \sqrt{2}
    \\[2mm]
    \CS(A_\alpha) = - \frac{1}{24} \mod 1
  \end{gathered}
\end{equation}
This result
with an asymptotic behavior~\eqref{E6_asymptotics}
coincides with an ansatz~\eqref{Z_asymptotics}.

\subsection{\mathversion{bold}
  $M(2,3,4)$
}

Let $\mathcal{E}_7$ be ${M}(2,3,4)$,
and we have
\begin{gather*}
  e(\mathcal{E}_7) = \frac{1}{12}
\end{gather*}
The  linking matrix~\eqref{Seifert_surgery}
can be transformed
by the Kirby move
to a form  of the Coxeter--Dynkin diagram
for an exceptional Lie algebra $E_7$;
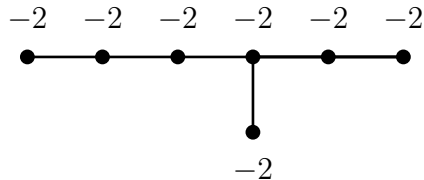
\begin{equation*}
  \raisebox{-1.6cm}{
    \begin{pspicture}(-4,-1.5)(6,1)
      \psdots[dotsize=4pt 2](-2,0)(-1,0)(0,0)(1,0)(2,0)(3,0)(1,-1)
      \pscustom[linewidth=1.0pt]{
        \psline{-}(-2,0)(-1,0)(0,0)(1,0)(2,0)(3,0)
        \psline{-}(1,0)(1,-1)
      }
      \rput{0}(-2,0.5){$-2$}
      \rput{0}(-1,0.5){$-2$}
      \rput{0}(0,0.5){$-2$}
      \rput{0}(1,0.5){$-2$}
      \rput{0}(2,0.5){$-2$}
      \rput{0}(3,0.5){$-2$}
      \rput{0}(1,-1.5){$-2$}
    \end{pspicture}
  }
\end{equation*}

\begin{prop}
  The WRT invariant for $\mathcal{E}_7$ is written in terms of a
  limiting value of the
  Eichler integrals
  $\widetilde{\Psi}_{12}^{(a)}(1/N)$
  as
  \begin{multline}
    \label{WRT_E7}
    \E^{\frac{37}{24 N} \pi \I} \,
    \left(
      \E^{\frac{2 \pi\I}{N}} - 1
    \right) \cdot
    \tau_N\left(
      \mathcal{E}_7
    \right)
    \\
    =
    \frac{\sqrt{2}}{4} \, \left( 1 + (-1)^N \right) \,
    \left(
      2 \, \E^{\frac{\pi \I}{24 N}}
      -
      \widetilde{\Psi}_{12}^{(1)}(1/N)
      -    \widetilde{\Psi}_{12}^{(5)}(1/N)
      -    \widetilde{\Psi}_{12}^{(7)}(1/N)
      -    \widetilde{\Psi}_{12}^{(11)}(1/N)
    \right)
  \end{multline}
\end{prop}

\begin{proof}
  The method is essentially same with  Prop.~\ref{prop:E6},
  although
  we have $\vv{p}=(2,3,4)$ and $\vv{q}=(-1,1,1)$ in this case.
  Using symmetries of the summand of~\eqref{Lawrence_Rozansky_form},
  the sum
  $\sum_{k_0=1}^{N-1} \sum_{n_j \mod p_j}$
  with $\vv{p}=(2,3,4)$ 
  can be replaced with a sum
  $\sum_{\substack{
      k_0 = a + 2 N n
      \\
      1 \leq a \leq N-1
      \\
      0 \leq n \leq 5
    }}
  \sum_{n_3 = 0}^3$ and   $n_1=n_2=0$.
  Taking a sum of $n_3$  and using a symmetry of the summand,
  we  obtain
  \begin{multline*}
    \text{l.h.s. of \eqref{WRT_E7}}
    =
    \frac{\E^{\frac{\pi \I}{4}}}{4 \, \sqrt{3 \, N}}
    \sum_{\substack{
        k_0 =0 \\
        N \nmid k_0
      }}^{12 N}
    \frac{
      \left(
        \E^{\frac{\pi \I}{2 N} k_0} -
        \E^{- \frac{\pi \I}{2 N} k_0}
      \right) \, 
      \left(
        \E^{\frac{\pi \I}{3 N} k_0} -
        \E^{- \frac{\pi \I}{3 N} k_0}
      \right)
    }{
      \E^{\frac{\pi \I}{ N} k_0} -
      \E^{- \frac{\pi \I}{ N} k_0}
    }
    \,
    \E^{ - \frac{\pi \I}{24 N} k_0^{~2}}
    \\
    \times
    \left(
      \E^{\frac{\pi \I}{4 N} k_0}
      \left(
        1 - \delta_{2 \mid k_0}
        +
        \E^{-\frac{\pi \I}{2} N} \,
        \left( -1 \right)^{\frac{k_0-1}{2}} \,
        \delta_{2 \mid (k_0 -1)}
      \right)
    \right.
    \\
    \left.
      -
      \E^{-\frac{\pi \I}{4 N} k_0}
      \left(
        1 - \delta_{2 \mid k_0}
        +
        \E^{-\frac{\pi \I}{2} N} \,
        \left( -1 \right)^{\frac{k_0  +1}{2}} \,
        \delta_{2 \mid (k_0 + 1)}
      \right)
    \right)
  \end{multline*}
  We introduce an infinitesimal variable $t$ in the fraction, and
  apply~\eqref{define_chi12}.
  We then get
  \begin{multline*}
    \text{l.h.s. of \eqref{WRT_E7}}
    =
%
    \frac{\E^{\frac{\pi \I}{4}}}{4 \, \sqrt{3 \, N}}
    \lim_{t \searrow 0} \sum_{n=0}^\infty
    \chi_{12}(n) \, \E^{-n t}
    \sum_{
        k =0 }^{6 N}
    \E^{\frac{\pi \I}{6 N} (2 k+1) n
      - \frac{\pi \I}{24 N} (2 k+1)^{2}}
    \\
    \times
    \left(
      \E^{\frac{\pi \I}{4 N} (2 k+1)}
      \left(
        1 
        +
        \E^{-\frac{\pi \I}{2} N} \,
        \left( -1 \right)^{k}
      \right)
      -
      \E^{-\frac{\pi \I}{4 N} (2 k+1)}
      \left(
        1
        +
        \E^{-\frac{\pi \I}{2} N} \,
        \left( -1 \right)^{k+1} 
      \right)
    \right)
  \end{multline*}
  In this computation, we need to subtract a sum over $N\mid k_0$, but 
  it is proved to vanish.
  We apply the Gauss reciprocity formula~\eqref{Gauss_reciprocity}, and
  then obtain
  \begin{align*}
    \text{l.h.s. of \eqref{WRT_E7}}
    & =
    \frac{\sqrt{2}}{4} \,
    \left(
      1+ \left(-1 \right)^N
    \right) \,
    \E^{\frac{3 \pi \I}{8 N}}
    \lim_{t \searrow 0} \sum_{n=0}^\infty
    \E^{- n \, t} \, \chi_{12}(n)     \,
    \E^{\frac{n^2}{6 N}\pi \I}
    \,
    \left(
      \E^{\frac{n}{2 N} \pi \I}
      -
      \E^{- \frac{n}{2 N} \pi \I}
    \right)
    \\
    & =
    \frac{\sqrt{2}}{4} \, \left( 1+ \left( -1 \right)^N \right)
    \,
    \sum_{k=1}^{12 N}
    \chi_{12}(k) \, 
    B_1 \left( \frac{k}{12 \, N} \right) \,
    \left(
      \E^{\frac{\pi \I}{6 N} \left( k + \frac{3}{2} \right)^2}
      -
      \E^{\frac{\pi \I}{6 N} \left( k - \frac{3}{2} \right)^2}
    \right)
  \end{align*}
  Finally we replace a sum of $k$ by $n=2 \, k \pm 3$.
  After some algebra, we obtain~\eqref{WRT_E7}.
\end{proof}

This result  proves   that the SU(2) WRT  invariant
$\tau_N(\mathcal{E}_7)$
vanishes when $N$ is odd.
Due to the factorization property~\eqref{factorize_tau}, this indicates that
\begin{equation}
  \tau_3
  \left(
    \mathcal{E}_7
  \right)=0
\end{equation}
which can be directly 
checked  from~\eqref{define_tau3} using 
the $E_7$ Dynkin matrix as
the
linking
matrix.

Asymptotic expansion of $\tau_N(\mathcal{E}_7)$ in $N\to\infty$
directly follows from~\eqref{WRT_E7} with a help
of~\eqref{Psi_nearly_modular}.

\begin{coro}
  Exact asymptotic expansion of the WRT invariant for $\mathcal{E}_7$
  in $N\to\infty$
  is 
  \begin{multline}
    \E^{\frac{37}{24 N} \pi \I} \,
    \left(
      \E^{\frac{2 \pi\I}{N}} - 1
    \right) \cdot
    \tau_N\left(
      \mathcal{E}_7
    \right)
    \simeq
    \frac{1 + (-1)^N}{\sqrt{2}}  \,
    \sqrt{\frac{N}{\I}} \, \E^{-\frac{1}{24} \pi \I N}
    +
    \frac{1 + (-1)^N}{\sqrt{2}} \,  \E^{\frac{\pi \I}{24 N}}
    \\
    -
    \frac{\sqrt{2}}{4} \,
    \left( 1 + (-1)^N \right) \,
    \sum_{k=0}^\infty
    \frac{
      L\left(
        - 2 \, k, 
        \psi_{24}^{(1)}+  \psi_{24}^{(5)}
        + \psi_{24}^{(7)}+  \psi_{24}^{(11)} 
      \right)
    }{
      k!}
    \left(
      \frac{\pi \,  \I}{24 \, N}
    \right)^k
  \end{multline}
  Here  the $L$-function is given by
  \begin{equation*}
    2 \,
    \frac{
      \ch(3 \,z ) \, \ch(2 \, z)
    }{
      \ch(6 \, z)
    }
    =
    \sum_{k=0}^\infty
    \frac{
      L \left(
        - 2 \, k ,
        \psi_{24}^{(1)} +    \psi_{24}^{(5)}+
        \psi_{24}^{(7)}+    \psi_{24}^{(11)}
      \right)
    }{
      (2 \, k) !
    } \,
    z^{2 k}
  \end{equation*}
  Then
  an asymptotic behavior of the partition function
  $Z_{N-2}(\mathcal{E}_7)$
  in $N\to\infty$ is
  \begin{equation}
    \label{E7_asymptotics}
    Z_{N-2}(\mathcal{E}_7)
    \sim
    \frac{1}{2} \, \E^{-\frac{3}{4} \pi \I} \,
    \left(
      \E^{-\frac{1}{24} \pi \I N}
      +
      \E^{-\frac{25}{24} \pi \I N}
    \right)
  \end{equation}
\end{coro}

By setting $\vv{p}=(2,3,4)$ and $\vv{q}=(-1,1,1)$
in~\eqref{torsion_ell} and~\eqref{CS_ell}, we obtain the torsion
and the Chern--Simons invariant as follows;
\begin{equation}
  \renewcommand{\arraystretch}{1.4}
  \begin{array}{c|cc}
    \vv{\ell} & \sqrt{T_\alpha} & \CS(A_\alpha)
    \\
    \hline
    (1,1,1) & 1 & -\frac{1}{48}
    \\
    (1,1,3) & 1 & -\frac{25}{48}
  \end{array}
\end{equation}
Substituting this result for~\eqref{Z_asymptotics},
we  recover~\eqref{E7_asymptotics}.

\subsection{\mathversion{bold}
  Poincar{\'e} Homology Sphere
  $M(2,3,5)$
}

Let $\mathcal{E}_8$ be ${M}( 2,3,5 )$, \emph{i.e.},
the Poincar{\'e} homology sphere, which has a following
$E_8$ Coxeter--Dynkin diagram as a  surgery
description;
\begin{equation*}
  \raisebox{-1.6cm}{
    \begin{pspicture}(-6,-2)(6,1)
      \psdots[dotsize=4pt 2](-3,0)(-2,0)(-1,0)(0,0)(1,0)(2,0)(3,0)(1,-1)
      \pscustom[linewidth=1.0pt]{
        \psline{-}(-3,0)(-2,0)(-1,0)(0,0)(1,0)(2,0)(3,0)
        \psline{-}(1,0)(1,-1)
      }
      \rput{0}(-3,0.5){$-2$}
      \rput{0}(-2,0.5){$-2$}
      \rput{0}(-1,0.5){$-2$}
      \rput{0}(0,0.5){$-2$}
      \rput{0}(1,0.5){$-2$}
      \rput{0}(2,0.5){$-2$}
      \rput{0}(3,0.5){$-2$}
      \rput{0}(1,-1.5){$-2$}
    \end{pspicture}
  }
\end{equation*}
The Euler characteristic is
\begin{gather*}
  e(\mathcal{E}_8) = \frac{1}{30}
\end{gather*}

As was demonstrated by Lawrence and Zagier, the WRT invariant for
$\mathcal{E}_8$ can be written in the following form;

\begin{prop}[\cite{LawrZagi99a},  also Ref.~\citen{KHikami04b}]
  The WRT invariant for the Poincar{\'e} homology sphere is written as
  \begin{multline}
    \label{WRT_E8}
    \E^{\frac{121}{60 N} \pi \I} \,
    \left(
      \E^{\frac{2 \pi\I}{N}} - 1
    \right) \cdot
    \tau_N\left( \mathcal{E}_8
    \right)
    \\
    =
    \E^{\frac{\pi \I}{60 N}}
    -
    \frac{1}{2} \,
    \left(
      \widetilde{\Psi}_{30}^{(1)}(1/N)
      +    \widetilde{\Psi}_{30}^{(11)}(1/N)
      +    \widetilde{\Psi}_{30}^{(19)}(1/N)
      +    \widetilde{\Psi}_{30}^{(29)}(1/N)
    \right)
  \end{multline}
\end{prop}

\begin{proof}
  We omit the proof. See Refs.~\citen{LawrZagi99a,KHikami04b}.
\end{proof}

Applying~\eqref{Psi_nearly_modular}, we obtain the asymptotic
expansion of the WRT invariant for the Poincar{\'e} homology sphere
in $N\to\infty$.

\begin{coro}[\cite{LawrZagi99a}]
  Exact asymptotic expansion of the WRT invariant for the Poincar{\'e}
  homology sphere in $N\to\infty$
  is 
  \begin{multline}
    \E^{\frac{121}{60 N} \pi \I} \,
    \left(
      \E^{\frac{2 \pi\I}{N}} - 1
    \right) \cdot
    \tau_N\left( \mathcal{E}_8
    \right)
    \\
    \simeq
    \sqrt{\frac{N}{\I}} \,
    \frac{2}{\sqrt{5}} \,
    \left(
      \sin \left(\frac{\pi}{5}\right) \, \E^{-\frac{1}{60} \pi \I N}
      +
      \sin \left(\frac{2 \, \pi}{5}\right) \, \E^{-\frac{49}{60} \pi \I N}
    \right)
    +
    \E^{\frac{\pi \I}{60 N}}
    \\
    +
    \frac{1}{2}
    \sum_{k=0}^\infty
    \frac{
      L\left(
        - 2 \, k, 
        - \psi_{60}^{(1)} -  \psi_{60}^{(11)}
        - \psi_{60}^{(19)} -  \psi_{60}^{(29)} 
      \right)
    }{
      k!}
    \left(
      \frac{\pi \,  \I}{60 \, N}
    \right)^k
  \end{multline}
  where we have the generating function of the $L$-function as
  \begin{equation*}
    2 \, \frac{
      \ch(5 \,z ) \, \ch(9 \, z)
    }{
      \ch(15 \, z)
    }
    =
    -
    \sum_{k=0}^\infty
    \frac{
      L \left(
        - 2 \, k ,
        -\psi_{60}^{(1)} -   \psi_{60}^{(11)}
        - \psi_{60}^{(19)} -    \psi_{60}^{(29)}
      \right)
    }{
      (2 \, k) !
    } \,
    z^{2 k}
  \end{equation*}
  Then we have an
  asymptotic behavior of the partition function
  $Z_{N-2}(\mathcal{E}_8)$
  in $N\to\infty$ as
  \begin{equation}
    \label{E8_asymptotics}
    Z_{N-2}(\mathcal{E}_8)
    \sim
    \frac{1}{2} \, \E^{-\frac{3}{4} \pi \I} \,
    \left(
      \sqrt{
        \frac{5 - \sqrt{5}}{5}
      } \,
      \E^{-\frac{1}{60} \pi \I N}
      +
      \sqrt{
        \frac{5 + \sqrt{5}}{5}
      } \,
      \E^{-\frac{49}{60} \pi \I N}
    \right)
  \end{equation}
\end{coro}

The torsion and the Chern--Simons invariant are given from~\eqref{torsion_ell} and~\eqref{CS_ell} by setting
$\vv{p}=(2,3,5)$ and $\vv{q}=(-1,1,1)$;
\begin{equation}
  \renewcommand{\arraystretch}{1.84}
  \begin{array}{c|cc}
    \vv{\ell} & \sqrt{T_\alpha} & \CS(A_\alpha)
    \\
    \hline
    (1,1,1) &
    2 \, \sqrt{\frac{2}{5}} \,  \sin\left(\frac{\pi}{5}\right)
    & -\frac{1}{120}
    \\
    (1,1,3) &
    2 \, \sqrt{\frac{2}{5}} \,  \sin\left(\frac{2 \, \pi}{5}\right)
    & -\frac{49}{120}
  \end{array}
\end{equation}
which supports~\eqref{Z_asymptotics}.

\subsection{\mathversion{bold}
  $M(2,2,K)$
}

Let $\mathcal{D}_{K}$ be the prism manifold ${M}(2,2,K)$
where we assume $K \geq 2$.
This manifold, which is defined by~\eqref{Seifert_surgery},
also has the following surgery description as of the Dynkin diagram
for $D_{K+2}$;
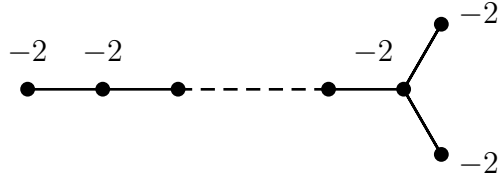
\begin{equation*}
  \raisebox{-1.6cm}{
    \begin{pspicture}(-6,-1.5)(6,1.5)
      \psdots[dotsize=4pt 2](-3,0)(-2,0)(-1,0)(1,0)(2,0)(2.5,0.866)(2.5,-0.866)
      \pscustom[linewidth=1.0pt]{
        \psline(-3,0)(-2,0)(-1,0)
      }
      \pscustom[linewidth=1.0pt]{
        \psline(1,0)(2,0)
        \psline(2.5,-0.866)(2,0)
        \psline(2.5,0.866)(2,0)
      }
      \psline[linewidth=1.0pt, linestyle=dashed]{-}(-1,0)(1,0)
      
      \rput{0}(-3,0.5){$-2$}
      \rput{0}(-2,0.5){$-2$}
      \rput{0}(1.6,0.5){$-2$}
      \rput{0}(3,1){$-2$}
      \rput{0}(3,-1){$-2$}
    \end{pspicture}
  }
\end{equation*}
where we have $K+2$ vertices $\bullet$.
Note that the Euler characteristic is
\begin{gather*}
  e(\mathcal{D}_K)=\frac{1}{K}
\end{gather*}

\begin{prop}
  The WRT invariant for $\mathcal{D}_{K}$ is written as a sum of the
  Eichler integrals, $\widetilde{\Psi}_{K}^{(1)}(1/N)$
  and
  $\widetilde{\Psi}_{K}^{(K-1)}(1/N)$,
  as follows.
  \begin{itemize}
  \item 
    $K$ is even;
    \begin{multline}
      \label{WRT_Deven}
      \E^{\frac{1}{2 N} \left(
          \sqrt{K} - \frac{1}{\sqrt{K}}
        \right)^2 \pi \I} \,
      \left(
        \E^{\frac{2 \pi\I}{N}} - 1
      \right) \cdot
      \tau_N\left( \mathcal{D}_{K}
      \right)
      \\
      =
      \left(
        1+ (-1)^{N \left(1 + \frac{ K}{2} \right)}
      \right) \,
      \left(
        \E^{\frac{\pi \I}{2 K N}}
        -
        \widetilde{\Psi}_{K}^{(1)}(1/N)
        -
        \widetilde{\Psi}_{K}^{(K-1)}(1/N)
      \right)
    \end{multline}

  \item $K$ is odd;
    \begin{multline}
      \label{WRT_Dodd}
      \E^{\frac{1}{2 N} \left(
          \sqrt{K} - \frac{1}{\sqrt{K}}
        \right)^2 \pi \I} \,
      \left(
        \E^{\frac{2 \pi\I}{N}} - 1
      \right) \cdot
      \tau_N\left( \mathcal{D}_{K}
      \right)
      \\
      =
      \left(
        1 + \E^{- \frac{K N}{2} \pi \I}
      \right) \,
      \E^{\frac{\pi \I}{2 K N}} 
      -
      \widetilde{\Psi}_{K}^{(1)}(1/N)
      -
      \widetilde{\Psi}_{K}^{(K-1)}(1/N)
      \\
      +
      \E^{- \frac{K N}{2} \pi \I} \,
      \left(
        -\widetilde{\Psi}_{K}^{(1)}(1/N)
        +
        \widetilde{\Psi}_{K}^{(K-1)}(1/N)
      \right)
    \end{multline}
  \end{itemize}
\end{prop}

\begin{proof}
  Using symmetries of the summand of~\eqref{Lawrence_Rozansky_form},
  the sum
  $\sum_{k_0=1}^{N-1} \sum_{n_j \mod p_j}$
  with $\vv{p}=(2,2,K)$
  can be replaced with a sum
  $\sum_{\substack{
      k_0 = a + 2 N n
      \\
      1 \leq a \leq N-1
      \\
      0 \leq n \leq K-1
    }}
  \sum_{n_1 = 0}^1
  \sum_{n_2 = 0}^1
  $ with   $n_3=0$.
  Taking sums over $n_1$ and $n_2$,
  and using a symmetry of the
  summand under $k_0 \to 2 \, N \, K - k_0$, we get
  \begin{multline*}
    \text{l.h.s. of \eqref{WRT_Deven}}
    \\
    =
    \frac{\E^{\frac{\pi \I}{4}}}{\sqrt{ 8 \, K \, N}}
    \sum_{\substack{
        k_0=0 \\
        N \nmid k_0
      }}^{2 K N}
    \frac{
      \left(
        \E^{\frac{\pi \I}{2 N} k_0} -
        \E^{- \frac{\pi \I}{2 N} k_0}
      \right) \, 
      \left(
        \E^{\frac{\pi \I}{K N} k_0} -
        \E^{- \frac{\pi \I}{K N} k_0}
      \right)
    }{
      \E^{\frac{\pi \I}{2 N} k_0} +
      \E^{- \frac{\pi \I}{2 N} k_0}
    }
    \,
    \E^{ - \frac{\pi \I}{2 K N} k_0^{~2}} \,
    \left(
      1-
      \left(-1 \right)^{N+k_0}
    \right)
  \end{multline*}

  We first assume that $K$ is even.
  In this case, we use the periodic function
  \begin{equation*}
    \renewcommand{\arraystretch}{1.4}
    \begin{array}{c|ccccc}
      n \mod 2 \, K & \frac{K}{2} -1  & \frac{K}{2}+1
      & \frac{3 \,  K}{2}-1 & \frac{3\, K}{2}+1 &  \text{others}
      \\
      \hline
      \varphi_{2 K}^{(e)}(n) & 1 & -1 & -1 & 1  & 0
    \end{array}
  \end{equation*}
  which satisfies
  \begin{equation}
    \frac{z - z^{-1}}{z^{\frac{K}{2}} + z^{- \frac{K}{2}}}
    =
    -
    \sum_{n=0}^\infty
    \varphi_{2K}^{(e)}(n) \, z^n
  \end{equation}
  We get
  \begin{multline*}
    \text{l.h.s. of \eqref{WRT_Deven}}
    \\
    =
    -\frac{\E^{\frac{\pi \I}{4}}}{\sqrt{ 8 \, K \, N}}
    \lim_{t \searrow 0}
    \sum_{n=0}^\infty \varphi_{2 K}^{(e)}(n) \, \E^{-n t}
    \sum_{\substack{
        k_0=0
        \\
        N \nmid k_0
      }}^{2 K N}
    \E^{ \frac{\pi \I}{N K} k_0 n - \frac{\pi \I}{2 K N} k_0^{~2}}
    \left(
      \E^{\frac{\pi \I}{2 N} k_0} -
      \E^{- \frac{\pi \I}{2 N} k_0} 
    \right) \,
    \left(
      1- (-1)^{N+k_0}
    \right)
  \end{multline*}
  We can check that the sum over $N \mid k_0$ vanishes.
  By applying
  the Gauss reciprocity formula~\eqref{Gauss_reciprocity}
  and taking a limit $t\searrow 0$,
  we get
  \begin{equation*}
    \text{l.h.s. of \eqref{WRT_Deven}}
    =
    \frac{1 + (-1)^{N + \frac{K}{2} N}}{2} \,
    \sum_{n=0}^{2 K N}
    \varphi_{2 K}^{(e)}(n) \, B_1 \left( \frac{n}{2 \, K \, N} \right)
    \,
    \left(
      \E^{\frac{\pi \I}{2 K N} \left( n+ \frac{K}{2} \right)^2}
      -
      \E^{\frac{\pi \I}{2 K N} \left( n- \frac{K}{2} \right)^2}
    \right)
  \end{equation*}
  As we have
  $\varphi_{2 K}^{(e)}\left( n \pm \frac{K}{2} \right) =
  \mp \left(
    \psi_{2 K}^{(1)}(n) + \psi_{2 K}^{(K-1)}(n)
  \right)$, we obtain~\eqref{WRT_Deven} after some manipulations.

  In the case that  $K$ is odd, we use the periodic function
  \begin{equation*}
    \begin{array}{c|ccccc}
      n \mod 4 \, K & K-2  & K+2
      & 3 \, K -2 & 3\, K+ 2  &  \text{others}
      \\
      \hline
      \varphi_{4 K}^{(o)}(n) & 1 & -1 & -1 & 1  & 0
    \end{array}
  \end{equation*}
  which has the following generating function;
  \begin{equation}
    \frac{z^2 - z^{-2}}{z^{K} + z^{- K}}
    =
    -
    \sum_{n=0}^\infty
    \varphi_{4 K}^{(o)}(n) \, z^n
  \end{equation}
  Using the same method, we obtain
  \begin{multline*}
    \text{l.h.s. of \eqref{WRT_Dodd}}
    =
    \frac{1}{2} \sum_{n=0}^{4 N K-1}
    \varphi_{4 K}^{(o)}(n) \, B_1 \left(\frac{n}{4 \, N \, K}\right)
    \\
    \times
    \left(
      \E^{\frac{\pi \I}{8 K N} (n+K)^2} \,
      \left(
        1 - (-1)^N \, \E^{\frac{n+K}{2} \pi \I + \frac{K N}{2} \pi \I}
      \right)
      -
      \E^{\frac{\pi \I}{8 K N} (n-K)^2} \,
      \left(
        1 - (-1)^N \, \E^{\frac{n-K}{2} \pi \I + \frac{K N}{2} \pi \I}
      \right)
    \right)
  \end{multline*}
  When we use  a relationship between $\varphi_{4K}^{(o)}(n \pm K)$
  and $\psi_{2 K}^{(a)}(n)$ as in the case of even $K$,
  we obtain~\eqref{WRT_Dodd}.
\end{proof}

This proposition indicates that the WRT invariant
$\tau_N(\mathcal{D}_{K})$ vanishes if $N$ is odd and $K \equiv 0
\mod 4$.
By use of~\eqref{define_tau3} with the Coxeter--Dynkin type
linking matrix,
we can check directly
\begin{equation}
  \tau_3 \left( \mathcal{D}_{K} \right)
  = 0
  \qquad
  \text{if $K \equiv 0 \mod 4$}
\end{equation}
The factorization property~\eqref{factorize_tau} proves this fact.

The exact asymptotic expansion of the WRT invariant
in $N\to\infty$
simply follows
from~\eqref{Psi_nearly_modular}.

\begin{coro}
  Exact asymptotic expansion of  the WRT invariant
  $\tau_N(\mathcal{D}_{K})$ in $N\to\infty$
  is given as follows.
  \begin{itemize}
  \item for even $K$;
    \begin{multline}
      \E^{\frac{1}{2 N} \left(
          \sqrt{K} - \frac{1}{\sqrt{K}}
        \right)^2 \pi \I} \,
      \left(
        \E^{\frac{2 \pi\I}{N}} - 1
      \right) \cdot
      \tau_N\left( \mathcal{D}_{K}
      \right)
      \simeq
      \left(1 + (-1)^{N+\frac{N K}{2}}\right)
      \,
      \Biggl(
      \sqrt{\frac{N}{\I}} \,
      \sqrt{\frac{2}{K}}
      \\
      \times
      \left(
        \sin\left(\frac{K}{4} \, \pi \right) \,
        \cos\left(\frac{K-2}{4} \, \pi\right)    \,
        \E^{-\frac{K}{8} \pi \I N}
        +
        2 \sum_{b=1}^{\frac{K}{2} -1}
        \sin\left(\frac{b}{2} \, \pi \right) \,
        \cos\left(\frac{K-2}{2 K} \, b \,  \pi \right) \,
        \E^{- \frac{b^2}{2 K} \pi \I N}
      \right)
      \\
      +
      \E^{\frac{\pi \I}{2 K N}}
      -
      \sum_{n=0}^\infty
      \frac{
        L\left(
          -2 \, n , \psi_{2 K}^{(1)} + \psi_{2 K}^{(K-1)}
        \right)
      }{n!} \,
      \left(
        \frac{\pi \, \I}{2 \, K \, N}
      \right)^n
      \Biggr)
    \end{multline}

  \item for odd $K$;
    \begin{multline}
      \E^{\frac{1}{2 N} \left(
          \sqrt{K} - \frac{1}{\sqrt{K}}
        \right)^2 \pi \I} \,
      \left(
        \E^{\frac{2 \pi\I}{N}} - 1
      \right) \cdot
      \tau_N\left( \mathcal{D}_{K}  \right)
      \\
      \simeq
      \sqrt{\frac{N}{\I}} \, 
      \sqrt{\frac{8}{K}} \,
      \sum_{b=1}^{\frac{K-1}{2}}
      \left(
        \sin\left(\frac{b}{2}  \, \pi \right) \,
        \cos\left(\frac{K-2}{2 K} \, b \,  \pi \right)
        -
        \E^{-\frac{K N}{2} \pi \I} \,
        \cos\left(\frac{b}{2} \, \pi \right) \,
        \sin\left(\frac{K-2}{2 K} \,  b \,  \pi \right)
      \right) \,
      \E^{-\frac{b^2}{2 K} \pi \I N}
      \\
      +
      \E^{\frac{\pi \I}{2 K N}} \,
      \left(
        1 + \E^{- \frac{K N}{2 } \pi \I}
      \right)
      \\
      -
      \sum_{n=0}^\infty
      \frac{
        1
      }{n!} \,
      \left(
        L\left(
          -2 \, n , \psi_{2 K}^{(1)} + \psi_{2 K}^{(K-1)}
        \right)
        -
        \E^{- \frac{K N}{2} \pi \I} \,
        L\left(
          -2 \, n ,  -  \psi_{2 K}^{(1)} + \psi_{2 K}^{(K-1)}
        \right)
      \right) \, 
      \left(
        \frac{\pi \, \I}{2 \, K \, N}
      \right)^n
    \end{multline}
  \end{itemize}
  Here the $L$-functions are computed from the generating functions as
  \begin{gather*}
    \frac{
      \ch \left( \frac{K-2}{2} \, z \right)
    }{
      \ch \left( \frac{K}{2}  \, z \right)
    }
    =
    \sum_{k=0}^\infty
    \frac{
      L \left(
        - 2 \, k ,  \psi_{2 K}^{(1)} + \psi_{2 K}^{(K-1)}
      \right)
    }{
      (2 \, k) !
    } \,
    z^{2 k}
    \\[2mm]
    \frac{
      \sh \left(\frac{K-2}{2} \, z \right)
    }{
      \sh \left( \frac{K}{2}  \, z \right)
    }
    =
    \sum_{k=0}^\infty
    \frac{
      L \left(
        - 2 \, k ,  \psi_{2 K}^{(1)} - \psi_{2 K}^{(K-1)}
      \right)
    }{
      (2 \, k) !
    } \,
    z^{2 k}
  \end{gather*}
  Thus a  dominating term of the partition function
  $Z_{N-2}(\mathcal{D}_{K})$
  in $N\to\infty$ is
  summarized by
  \begin{equation}
    \label{D_asymptotics}
    Z_{N-2}(\mathcal{D}_{K})
    \sim
    \frac{1}{2} \, \E^{-\frac{3}{4} \pi \I} \,
    \sum_{m=0}^{
      \left\lfloor \frac{K}{2} \right\rfloor -1
    }
    \frac{4}{\sqrt{K}} \,
    \sin \left( \frac{2 \, m+1}{K} \, \pi \right) \,
    \E^{
      - \frac{ (2 m+1)^2}{2 K} \pi \I N
    }
  \end{equation}
\end{coro}

For the manifold $\mathcal{D}_{K}$,
the torsion~\eqref{torsion_ell} and the Chern--Simons
invariant~\eqref{CS_ell} can be computed by setting 
$\vv{p}=(2,2,K)$ and $\vv{q}=(-1,1,1)$.
When we choose
$\vv{\ell}=(1,1, 2 \, m+1)$ with
$0 \leq m < \frac{K-1}{2}$, we obtain
\begin{equation}
  \begin{gathered}
    \sqrt{T_\alpha}
    =
    \frac{4}{\sqrt{K}} \,
    \left|
      \sin \left( \frac{2 \, m +1}{K} \, \pi \right)
    \right|
    \\[2mm]
    \CS(A_\alpha)
    =
    - \frac{\left(2 \, m+1\right)^2}{4 \, K}
    \mod 1
  \end{gathered}
\end{equation}
which supports~\eqref{Z_asymptotics}.

\subsection{Comments}
To close this section, we shall give several relations among the SU(2)
quantum invariants for the 3-manifolds and links.
First of all
we see that the WRT invariants for manifolds
$\mathcal{E}_6$ and $\mathcal{D}_6$ are related to each other, and
we have
\begin{equation}
  \E^{\frac{\pi \I}{N}} \cdot
  \tau_N \left( \mathcal{D}_6 \right)
  +
  \frac{
    \E^{- \frac{\pi \I}{N}}
  }{
    \E^{\frac{2 \pi \I}{N}} - 1
  }
  =
  \frac{4}{\sqrt{3}} \, \tau_N \left(\mathcal{E}_6\right)
  \qquad
  \text{
    for $3 \mid N$}
\end{equation}

Recalling a result~\eqref{Kashaev_link}
in Ref.~\citen{KHikami03a}, the WRT invariant for
$\mathcal{D}_2$ is related to  Kashaev's invariant for the torus link
$\mathcal{T}_{2,4}$;
\begin{equation}
  \label{D2_T24}
  \left(
    \E^{\frac{2 \pi \I}{N}} - 1
  \right) \,
  \tau_N
  \left(\mathcal{D}_2 \right)
  =
  2 \,
  \left(
    1 - \frac{1}{N} \,
    \left\langle \mathcal{T}_{2,4} \right\rangle_N
  \right)
\end{equation}
Topologically
this coincidence may be explained from a fact that
the Seifert manifold
$M(0; (k,-1),(k,1),(k,1))$ is constructed from 0-framed surgery of the torus
link $\mathcal{T}_{2,2k}$,
and
the manifold $\mathcal{D}_2$ is given from $\mathcal{T}_{2,4}$;
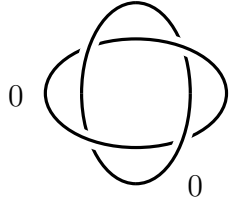
\begin{equation*}
  \begin{psfrags}
    \psfrag{0}{$0$}
    \includegraphics[scale=0.6, bb=0 -64 10 64]{knot24.ps}
  \end{psfrags}
\end{equation*}
Furthermore we  find that
the WRT invariant for $\mathcal{D}_K$ with odd $K$ is related to 
Kashaev's invariant for torus link $\mathcal{T}_{2,2K}$ as
\begin{equation}
  \label{D_odd_link}
  \left(
    \E^{\frac{2 \pi \I}{N}} - 1 
  \right) \,
  \tau_N  \left( \mathcal{D}_{K} \right)
  =
  - \frac{2}{K \, N} \,
  \left\langle \mathcal{T}_{2, 2 K} \right\rangle_N
  \qquad
  \text{for odd $K$ and $N \equiv 2 \mod 4$}
\end{equation}
We do not know a precise meaning of this relation,
but
a  connection between $\mathcal{D}_K$ and
$\mathcal{T}_{2,2K}$ might be explained  as follows.
The triangle group $T_{2,2,K}$ is a subgroup of $T_{2,2, 2K}$,
and the manifold $\mathcal{D}_{2K}$ is homeomorphic to the 2-fold
cyclic branched covering of $S^3$, branched along a torus link
$\mathcal{T}_{2,2K}$~\cite{JMiln75a}.



\section{Relationship with the Platonic Solids}
\label{sec:Platonic}

We have clarified that the  WRT invariant for the
spherical Seifert manifolds
$S^3/\Gamma$ with the finite subgroup $\Gamma$ of SU(2) is written in
terms of a limiting value  of the Eichler integrals of the half-integral
weight modular forms.
As the fundamental group~\eqref{fundamental_solid}
of these manifolds is related to the polyhedral group
as in Table~\ref{tab:spherical}
and
the manifold is a spherical neighborhood of the
Kleinian singularities associated to hypersurface in Table~\ref{tab:surface},
one may expect that the modular forms, whose Eichler
integrals denote  the WRT  invariant of the manifolds, have a
connection with the Platonic solid.
This type of relationship was conjectured in Ref.~\citen{GuadPilo98a},
and established was the connection
between the absolute value of the WRT invariant and  the fundamental
group for
a case of lens space~\cite{SYamad95a}.

Here we shall demonstrate this connection for several cases.
This may be compared with 
the ADE
classification of the modular invariant partition function of
the  conformal field theory~\cite{CappItzyZube87}
and the classification of the rational conformal field theories based
on the Fuchsian differential equation~\cite{MatMukSen88a,EBKirit89a}.

\subsection{
  Tetrahedral Group}

Our result~\eqref{WRT_E6} indicates that the WRT invariant
for $\mathcal{E}_6=M(2,3,3)$
is regarded as a
sum of the
Eichler integral of 
modular forms
$\Psi_6^{(1)}(\tau) + \Psi_6^{(5)}(\tau) $
and
$\Psi_6^{(3)}(\tau) $.
These two $q$-series span a two-dimensional vector modular form with
weight $3/2$, and when we set
\begin{align}
  \boldsymbol{\Psi}_{E_6}(\tau)
  & =
  \frac{1}{
    \left(
      \eta(\tau)
    \right)^3
  } \,
  \begin{pmatrix}
    \frac{1}{\sqrt{2}} \left(
      \Psi_6^{(1)}(\tau)  +       \Psi_6^{(5)}(\tau) 
    \right)
    \\[2mm]
    \Psi_6^{(3)}(\tau) 
  \end{pmatrix}
  \\
  & \equiv \nonumber
  \begin{pmatrix}
    X(\tau) \\[2mm]
    Y(\tau)
  \end{pmatrix}
  =
  \begin{pmatrix}
    \frac{1}{\sqrt{2}} \, q^{- \frac{1}{12}} \,
    \left(
      1 + 8 \, q + 17 \, q^2 + 46 \, q^3 + \cdots
    \right)
    \\[4mm]
    3 \, q^{\frac{1}{4}} \,
    \left(
      1 + 3 \, q + 9 \, q^2 + 19 \, q^3 + \cdots
    \right)
  \end{pmatrix}
\end{align}
we have the transformation property by use of~\eqref{Psi_modular}
\begin{equation}
  \label{ST_E6}
  \begin{gathered}
    \boldsymbol{\Psi}_{E_6}(\tau)
    =
    \frac{1}{\sqrt{3}} \,
    \begin{pmatrix}
      1 & \sqrt{2} \\[2mm]
      \sqrt{2} & -1
    \end{pmatrix} \,
    \boldsymbol{\Psi}_{E_6}(-1/\tau)
    \\[2mm]
    \boldsymbol{\Psi}_{E_6}(\tau+1)
    =
    \begin{pmatrix}
      \E^{- \frac{1}{6} \pi \I} & 0 \\[2mm]
      0 & \E^{\frac{1}{2} \pi \I}
    \end{pmatrix} \,
    \boldsymbol{\Psi}_{E_6}(\tau)
  \end{gathered}
\end{equation}
We note that we have divided the vector modular form
by powers of Dedekind $\eta$-function for
our physical convention, and
that this is the character for $k=1$ SU(3) WZW model
up to constant which can
be checked from the Verlinde formula~\cite{Verli88}.

We now look for invariant polynomials of $X$ and $Y$ under the modular
group~\eqref{ST_E6}.
As homogeneous polynomials of $X$ and $Y$, we define three polynomials
by~\cite{Ebeli94Book}
\begin{equation}
  \label{polynomial_E6}
  \begin{gathered}[h]
    V_T (X, Y) = X^4 + 2 \, \sqrt{2} \, X \, Y^3
    \\[2mm]
    F_T (X, Y)= Y^4 - 2 \, \sqrt{2} \, X^3 \, Y
    \\[2mm]
    E_T (X, Y)
    =X^6 - 5 \sqrt{2} \, X^3 \, Y^3 - Y^6
  \end{gathered}
\end{equation}
Based on the transformation laws~\eqref{ST_E6}, we can check directly 
that these
polynomials are invariant under the modular group.
Furthermore  the modular transformation property
shows that  they can be written as
\begin{gather}
  \begin{gathered}
    \left( \eta(\tau) \right)^8 \cdot
    V_T(\tau)
    = \frac{1}{4} \, E_4(\tau)
    \\[2mm]
    F_T(\tau)
    =
    -3
    \\[2mm]
%
    \left(\eta(\tau) \right)^{12} \cdot E_T(\tau)
    =
    \frac{1}{8} \, E_6(\tau)
  \end{gathered}
\end{gather}
where we have used the Eisenstein
series~\eqref{Eisenstein_series} and the Dedekind
$\eta$-function~\eqref{Dedekind_eta}.
By definition of the invariant polynomials~\eqref{polynomial_E6},
those  3 polynomials
satisfy the tetrahedral equation~\cite{Klein56Book}
\begin{equation}
  V_{T}^{~3} + F_T^{~3} = E_T^{~2}
\end{equation}
which reduces to
\begin{equation*}
  R(x,y,z) =
  x ^3 + y^4 + z^2 = 0
\end{equation*}
in Table~\ref{tab:surface} by setting
$x= - 4^{1/3} \, V_T \, F_T$,
$y= E_T$,
and
$z=\I \, \left( V_T^{~3} - F_T^{~3} \right)$.

The symmetry group of the tetrahedron can be derived by considering
the principal congruence subgroup  of $SL(2;\mathbb{Z})$.
We note
that
the  vector modular form $\boldsymbol{\Psi}_{E_6}(\tau)$
is  written as
the theta series on the root lattice of $A_2$;
\begin{equation}
  \left( \eta(\tau) \right)^2 \cdot
  \boldsymbol{\Psi}_{E_6}(\tau)
  =
  {
    \begin{pmatrix}
      {\displaystyle
        \frac{1}{\sqrt{2}}
        \sum_{(x,y)\in\mathbb{Z}^2}
        q^{x^2 - x y + y^2}
      }
      \\[8mm]
      {\displaystyle
        q^{\frac{1}{3}}
        \sum_{(x,y)\in\mathbb{Z}^2}
        q^{x^2 - x y + y^2+x - y}
      }
    \end{pmatrix}
  }
\end{equation}
We see from the transformation law of the right hand side that
it is a modular form with weight 1 for the group $\Gamma(3)$,
where  $\Gamma(M)$ 
is the principal congruence subgroup of level $M$ of
$SL(2;\mathbb{Z})$ 
(see, \emph{e.g.}, Refs.~\citen{Koblitz93Book,Ebeli94Book,DZagier92a})
\begin{equation}
  \Gamma(M)
  =
  \left\{
    \begin{pmatrix}
      a & b
      \\
      c & d
    \end{pmatrix}
    \in SL(2;\mathbb{Z})
    ~
    \Big|
    ~
    \begin{array}{l}
      a \equiv d \equiv 1 \mod M
      \\[2mm]
      b \equiv c \equiv 0 \mod M
    \end{array}
  \right\}
\end{equation}
This fact is based on that the level of the root lattice $A_2$ is 3.
The group $PSL(2;\mathbb{Z})/\Gamma(3)$  is isomorphic to
the symmetry group of the
tetrahedron~\cite{Ebeli94Book}.
Then we have a mapping
\begin{equation*}
  \boldsymbol{\Psi}_{E_6}
  : 
  \overline{\mathbb{H}/\Gamma(3)}
  \rightarrow \mathbb{P}^1
\end{equation*}
where $  \overline{\mathbb{H}/\Gamma(3)}$ means a compactification of
$\mathbb{H}/\Gamma(3)$ by adding a point $\infty$,
and the tetrahedral group  acts on the tetrahedron in the Riemann
sphere $\mathbb{P}^1$.

This action can be seen immediately by studying  the zeros of
the homogeneous polynomials, $V_T$, $F_T$, and $E_T$~\cite{Ebeli94Book}.
We consider the regular tetrahedron in $\mathbb{R}^3$, which is
inscribed in the unit sphere $S^2$ around the origin with the south
pole $(0,0,-1)$
as one of
vertices.
We  identify $S^2$ with 
$\mathbb{P}^1=\mathbb{C}\cup \{ \infty \}$
by the projection from the north pole $(0,0,1)$
to the equatorial plane,
and
we regard
$(X : Y)$  as the homogeneous coordinates of
$\mathbb{P}^1$ identifying $(1:0)$ with $\infty$.
Then we see that the four  zeros of the homogeneous polynomial
$V_T$~\eqref{polynomial_E6}
denote the vertices of the regular tetrahedron.
In the same manner, the zeros of the polynomial $F_T$ are the
mid-points of faces,
\emph{i.e.}, the
intersections of the unit sphere $S^2$ and the straight line which
connects the origin and each vertex of the tetrahedron.
The zeros of the homogeneous polynomial $E_T$ denote
the mid-points of edges,
\emph{i.e.},
the intersections of the sphere and the straight line which  connects
the middle point of edges which do not share the vertex of the tetrahedron.
As a consequence the polynomials $V_T$, $F_T$, and $V_T$
are  invariant under the
tetrahedral group.

\subsection{Octahedral Group}

We see from~\eqref{WRT_E7}
that the WRT invariant for $\mathcal{E}_7 = M(2,3,4)$
can be regarded as the Eichler integrals
of the $q$-series
$\Psi_{12}^{(1)}(\tau)  +       \Psi_{12}^{(5)}(\tau) 
      +      \Psi_{12}^{(7)}(\tau)  +       \Psi_{12}^{(11)}(\tau) 
$.
This function with 2 more functions
spans a 3-dimensional space of the modular form with
weight $3/2$;
when we define
the vector modular form $\boldsymbol{\Psi}_{E_7}(\tau)$ by
\begin{align}
  \boldsymbol{\Psi}_{E_7}(\tau)
  & =
  \frac{1}{
    \left( \eta(\tau) \right)^3
  } \,
  \begin{pmatrix}
    \frac{1}{2} \left(
      \Psi_{12}^{(1)}(\tau)  +       \Psi_{12}^{(5)}(\tau) 
      +      \Psi_{12}^{(7)}(\tau)  +       \Psi_{12}^{(11)}(\tau) 
    \right)
    \\[2mm]
    \frac{1}{\sqrt{2}}\left(
      \Psi_{12}^{(4)}(\tau) +
      \Psi_{12}^{(8)}(\tau) 
    \right)
    \\[2mm]
    \frac{1}{2} \left(
      \Psi_{12}^{(1)}(\tau)  -       \Psi_{12}^{(5)}(\tau) 
      +      \Psi_{12}^{(7)}(\tau)  -       \Psi_{12}^{(11)}(\tau) 
    \right)
  \end{pmatrix}
  \label{Psi_E7}
  \\
  & \equiv \nonumber
  \begin{pmatrix}
    X(\tau)
    \\[2mm]
    Y(\tau)
    \\[2mm]
    Z(\tau)
  \end{pmatrix}
  =
  \begin{pmatrix}
    \frac{1}{2} \, q^{-\frac{5}{48}} \,
    \left(
      1 + 5 \, q^{\frac{1}{2}} + 10 \, q
      + 15 \, q^{\frac{3}{2}} + \cdots
    \right)
    \\[4mm]
    2 \sqrt{2}  \, q^{\frac{5}{24}} \,
    \left(
      1 + 5 \, q + 15 \, q^2 + 40 \,  q^3 + \cdots
    \right)
    \\[4mm]
    \frac{1}{2} \, q^{-\frac{5}{48}} \,
    \left(
      1 - 5 \, q^{\frac{1}{2}} + 10 \, q
      - 15 \, q^{\frac{3}{2}} + \cdots
    \right)
  \end{pmatrix}
\end{align}
the modular transformation~\eqref{Psi_modular} reduces to
\begin{equation}
  \label{ST_E7}
  \begin{gathered}
    \boldsymbol{\Psi}_{E_7}(\tau)
    =
    \begin{pmatrix}
      1 & & \\[2mm]
      & 0 & 1 \\[2mm]
      & 1 & 0
    \end{pmatrix} \,
    \boldsymbol{\Psi}_{E_7}(-1/\tau)
    \\[2mm]
    \boldsymbol{\Psi}_{E_7}(\tau+1)
    =
    \begin{pmatrix}
      & & \E^{- \frac{5}{24} \pi \I} \\[2mm]
      & \E^{\frac{5}{12} \pi \I} & \\[2mm]
      \E^{- \frac{5}{24} \pi \I} & &
    \end{pmatrix} \,
    \boldsymbol{\Psi}_{E_7}(\tau)
  \end{gathered}
\end{equation}

We consider the homogeneous polynomials which are invariant under the
modular group~\eqref{ST_E7}.
Empirically we define polynomials as follows;
\begin{equation}
  \label{E7_polynomial}
  \begin{gathered}
    V_C (X, Y, Z) = \left( X \, Y \, Z  \right)^2
    \\[2mm]
    F_C (X, Y, Z) = X^8 - Y^8 - Z^8
    \\[2mm]
    \begin{aligned}
      \left(
        E_C (X, Y, Z)
      \right)^{2}
      & =
      X^{24} - Y^{24} - Z^{24} - \frac{3351}{4} \, (X \, Y \, Z)^8
      \\
      & \qquad
      -
      3 \,\left( Y^8+ Z^8 \right) \,
      \left( X^{16} +  Y^8 \, Z^8 \right)
      +
      3 \,\left( Y^{16}+ Z^{16} \right) \, X^{8}
    \end{aligned}
  \end{gathered}
\end{equation}
Using  modular transformation laws~\eqref{ST_E7} and
recalling properties of the
space of the modular form, we find that these invariant polynomials
can be written as
\begin{gather}
  \begin{gathered}
    V_C(\tau)
    =
    \frac{1}{2} 
    \\[2mm]
    \left( \eta(\tau) \right)^8 \,
    F_C(\tau)
    =
    \frac{5}{16} \, E_4(\tau)
    \\[2mm]
    \left( \eta(\tau) \right)^{12}
    \,
    E_C(\tau)
    =
    \frac{5 \, \sqrt{5}}{64} \, E_6(\tau)
  \end{gathered}
\end{gather}
The invariant polynomials $V_C$, $F_C$, and $E_C$ satisfy the cubic
equation~\cite{Klein56Book}
\begin{equation}
  E_C^{~2}=  F_C^{~3} - \frac{3375}{4} \, V_C^{~4}
\end{equation}
which follows from the definition~\eqref{E7_polynomial}, and
coincides with the identity~\eqref{E_E_Delta}.
After setting
$x=\frac{15^{3/2}}{2} \, V_C^{~2}$,
$y= -F_C$,
and $z=\frac{15^{3/4}}{\sqrt{2}} \, E_C \, V_C$,
we recover the  hypersurface for $E_7$;
\begin{equation*}
  R(x,y,z)
  =
  x^3 + x \, y^3 + z^2 = 0
\end{equation*}

To discuss the  modular group,
we recall the Jacobi theta functions
\begin{equation}
  \label{Jacobi_theta}
  \begin{gathered}
    \theta_{00}(\tau)
    =
    \sum_{n\in\mathbb{Z}} q^{\frac{1}{2} n^2}
    =
    \frac{
      \left(
        \eta \left(\frac{\tau+1}{2} \right)
      \right)^2
    }{
      \eta(\tau)
    }
    \\[2mm]
    \theta_{10}(\tau)
    =
    \sum_{n\in\mathbb{Z}} q^{\frac{1}{2} \left(n
        +\frac{1}{2}\right)^2}
    =
    2 \,
    \frac{
      \left( \eta(2 \, \tau) \right)^2
    }{
      \eta(\tau)
    }
    \\[2mm]
    \theta_{01}(\tau)
    =
    \sum_{n\in\mathbb{Z}} (-1)^n \, q^{\frac{1}{2} n^2}
    =
    \frac{
      \left( \eta \left(\frac{\tau}{2}\right) \right)^2
    }{
      \eta(\tau)
    }
  \end{gathered}
\end{equation}
When we set the vector  as
\begin{equation}
  \boldsymbol{\Theta}(\tau)
  =
  \begin{pmatrix}
    \theta_{00}(\tau)
    \\[2mm]
    \theta_{10}(\tau)
    \\[2mm]
    \theta_{01}(\tau)
  \end{pmatrix}
\end{equation}
this becomes a vector modular form with weight
$1/2$ (see, \emph{e.g.}, Ref.~\citen{Mumf83});
\begin{equation}
  \label{theta_modular}
  \begin{gathered}
    \boldsymbol{\Theta}(\tau)
    =
    \sqrt{\frac{\I}{\tau}} \cdot
    \begin{pmatrix}
      1 & &
      \\[2mm]
      & 0 & 1
      \\[2mm]
      & 1 & 0
    \end{pmatrix}
    \,
    \boldsymbol{\Theta}(-1/\tau)
    \\[2mm]
    \boldsymbol{\Theta}(\tau+1)
    =
    \begin{pmatrix}
      & & 1 \\[2mm]
      & \E^{\frac{1}{4} \pi \I} &
      \\[2mm]
      1 & &
    \end{pmatrix}
    \,
    \boldsymbol{\Theta}(\tau)
  \end{gathered}
\end{equation}
With these modular transformation formulae, we find that the functions
$X$, $Y$, and $Z$ in~\eqref{Psi_E7}, are written in terms of the Jacobi theta functions
\begin{equation}
  \begin{pmatrix}
    X^2 \\[2mm]
    Y^2 \\[2mm]
    Z^2
  \end{pmatrix}
  =
  \frac{1}{
    4 \, 
    \left( \eta(\tau) \right)^5}
  \begin{pmatrix}
    \left( \theta_{00} \right)^5
    \\[2mm]
    \left( \theta_{10} \right)^5
    \\[2mm]
    \left( \theta_{01} \right)^5
  \end{pmatrix}
\end{equation}
The transformation properties~\eqref{theta_modular} show that
the theta functions defined by
$
\boldsymbol{\Theta}^2(\tau)
=
\left(
  \left( \theta_{00} \right)^2 ,
  \left( \theta_{10} \right)^2 ,
  \left( \theta_{01} \right)^2 
\right)^t
$
is a modular form with weight $1$ for the group $\Gamma(4)$.
We then
have a map
\begin{equation*}
  \left(
    \boldsymbol{\Psi}_{E_7}
  \right)^{4} :
  \overline{\mathbb{H}/\Gamma(4)}
  \rightarrow
  \mathbb{P}^2
\end{equation*}
and the group
$PSL(2;\mathbb{Z})/\Gamma(4)$ denotes the symmetry of the cube.

Simple explanation of connection with the octahedral group is as follows.
We reconsider the modular group acting on 
$(x, y, z)
\equiv
\left(
  X^{24}, Y^{24}, Z^{24}
\right)
$.
Under the action of $S$ and $R=T \, S$, we have
\begin{align*}
  S & :
  \begin{pmatrix}
    x \\
    y \\
    z
  \end{pmatrix}
  \to
  \begin{pmatrix}
    x \\
    z \\
    y
  \end{pmatrix}
  &
  R = T \, S & :
  \begin{pmatrix}
    x \\
    y \\
    z
  \end{pmatrix}
  \to
  \begin{pmatrix}
    -z \\
    -x \\
    y
  \end{pmatrix}
\end{align*}
When we  interpret these actions on $\mathbb{R}^3$ with coordinates,
$x$, $y$, and $z$,
the action $S$ can be regarded as a reflection  at the plane 
$y=z$.
In the same way, the action $R$ denotes a $\frac{2 \pi}{3}$ rotation
around an axis which passes both the origin and $(-1, 1, 1)$.
As a result  the cube  whose vertices are on
$(\varepsilon_1, \varepsilon_2, \varepsilon_3)$ with
$\varepsilon_i=\pm 1$ is invariant under the modular group.

\subsection{Icosahedral Group}

We have seen that
the WRT invariant~\eqref{WRT_E8}
for the Poincar{\'e} homology sphere
$\mathcal{E}_8=M(2,3,5)$
is regarded as  the Eichler integral of
$        \Psi_{30}^{(1)}(\tau)  +    \Psi_{30}^{(11)}(\tau) 
    +      \Psi_{30}^{(19)}(\tau) +      \Psi_{30}^{(29)}(\tau) 
$
of weight $3/2$.
As was pointed out in Ref.~\citen{LawrZagi99a},
it spans a
two-dimensional vector modular form;
when we define
\begin{align}
  \boldsymbol{\Psi}_{E_8}(\tau)
  & =
  \frac{1}{
    \left(
      \eta(\tau)
    \right)^3} \,
  \begin{pmatrix}
    \Psi_{30}^{(1)}(\tau)  +    \Psi_{30}^{(11)}(\tau) 
    +      \Psi_{30}^{(19)}(\tau) +      \Psi_{30}^{(29)}(\tau) 
    \\[2mm]
    \Psi_{30}^{(7)}(\tau)  +    \Psi_{30}^{(13)}(\tau) 
    +      \Psi_{30}^{(17)}(\tau) +      \Psi_{30}^{(23)}(\tau) 
  \end{pmatrix}
  \\
  & \equiv \nonumber
  \begin{pmatrix}
    X(\tau)
    \\[2mm]
    Y(\tau)
  \end{pmatrix}
  =
  \begin{pmatrix}
    q^{-\frac{7}{60}} \,
    \left(
      1+ 14 \, q + 42 \, q^2 + 140 \, q^3 + \cdots
    \right)
    \\[4mm]
    q^{\frac{17}{60}} \,
    \left(
      7 + 34 \, q + 119 \, q^2 + 322 \, q^3 + \cdots
    \right)
  \end{pmatrix}
\end{align}
we have under the $S$- and $T$-transformations
\begin{equation}
  \label{E8_modular}
  \begin{gathered}
    \boldsymbol{\Psi}_{E_8}(\tau)
    =
    \frac{2}{\sqrt{5}} \,
    \begin{pmatrix}
      \sin \left(\frac{\pi}{5}\right) &
      \sin \left(\frac{2 \pi}{5}\right) 
      \\[2mm]
      \sin \left(\frac{2 \pi}{5}\right) &
      -    \sin \left(\frac{\pi}{5}\right) 
    \end{pmatrix} \,
    \boldsymbol{\Psi}_{E_8}(-1/\tau)
    \\[2mm]
    \boldsymbol{\Psi}_{E_8}(\tau+1)
    =
    \begin{pmatrix}
      \E^{-\frac{7}{30} \pi \I}  &\\
      & \E^{\frac{17}{30} \pi \I} 
    \end{pmatrix} \,
    \boldsymbol{\Psi}_{E_8}(\tau)
  \end{gathered}
\end{equation}
We notice that 
in the vector modular form $\boldsymbol{\Psi}_{E_8}(\tau)$
the  subscript $30$
is the Coxeter number of the Lie algebra $E_8$, and that a set of superscripts,
$\{1, 7, 11, 13, 17, 19, 23, 29 \}$,
also coincides with the exponents of $E_8$
(see, \emph{e.g.}, Ref.~\citen{Humph72Book}).
So  we may expect  that the vector modular form
$\boldsymbol{\Psi}_{E_8}(\tau)$ has a connection with the
exceptional Lie algebra $E_8$.
Although,
we note that the modular form $\boldsymbol{\Psi}_{E_8}(\tau)$ denotes the
character of the $k=1$ $G_2$ WZW model~\cite{Verli88,MatMukSen88a}.

To  find a more explicit and geometrical  relationship with the $E_8$
algebra,
we define three homogeneous  polynomials of $X$ and $Y$ 
following Klein~\cite{Klein56Book}
by
\begin{gather}
  \label{icosahedron_Klein_polynomial}
  \begin{gathered}
    V_I(X,Y)
    = X \, Y \, \left(
      X^{10} + 11 \, X^5 \, Y^5 - Y^{10}
    \right)
    \\[2mm]
    F_I(X,Y)
    =
    X^{20} + Y^{20} - 228 \, X^5 \, Y^5 \,
    \left( X^{10} - Y^{10} \right)
    +494 \, X^{10} \, Y^{10}
    \\[2mm]
    E_I(X,Y)
    =
    X^{30} + Y^{30} + 522 \, X^5 \, Y^5 \,
    \left(
      X^{20} - Y^{20}
    \right)
    -
    10005 \, X^{10} \, Y^{10} \,
    \left(
      X^{10} + Y^{10}
    \right)
  \end{gathered}
\end{gather}
We can check that these are invariant polynomials under the modular
group~\eqref{E8_modular}, and
by investigating the modular properties  these polynomials
are written in terms of the Eisenstein  series as
\begin{gather}
  \label{E8_polynomial}
  \begin{gathered}
    27 \,
    \Delta(\tau) \,
    V_I
    =
    125 \, \left( E_4 \right)^3 + 64 \,
    \left( E_6 \right)^2
    \\[2mm]
    2916 \,
    \left(
      \eta(\tau) \right)^{56} \,
    F_I
    =
    E_4 \, 
    \left(
      -3125 \, \left( E_4 \right)^6
      + 9625 \, \left( E_4 \right)^3 \, 
      \left( E_6 \right)^2
      -3584 \, 
      \left( E_6 \right)^4
    \right)
    \\[2mm]
    \begin{aligned}
      & 157464 \,
      \left( \eta(\tau)\right)^{84} \,
      E_I
      \\
      & =
      E_6 \,
      \left(
        546875 \, \left( E_4 \right)^9
        -931875 \,
        \left( E_4 \right)^6  \left( E_6 \right)^2
        +575232 \,
        \left( E_4 \right)^3  \left( E_6 \right)^4
        -32768 \,
        \left( E_6 \right)^6
      \right)
    \end{aligned}
  \end{gathered}
\end{gather}
We see by definition~\eqref{icosahedron_Klein_polynomial}
that these invariant polynomials satisfy the icosahedron equation
\begin{equation}
  1728 \, V_I^{~5} + F_I^{~3} = E_I^{~2}
\end{equation}
which reduces to the hypersurface for $E_8$,
\begin{equation*}
  R(x,y,z)= x^3 + y^5 + z^2 =0
\end{equation*}
when we set
$x=-F_I$,
$y=- 12^{1/5} \, V_I$,
and
$z=E_I$.

The modular transformation property~\eqref{E8_modular}  proves that
the  modular form
$\left( \eta(\tau) \right)^{14/5} \cdot \boldsymbol{\Psi}_{E_8}(\tau)$
with rational weight $7/5$
is on the group $\Gamma(5)$.
As was studied in Ref.~\citen{TIbuki00a}, the polynomial ring of
the group $\Gamma(5)$ is known to be spanned by
modular forms
$
\left( \eta(\tau) \right)^{2/5} \,
  \Phi_1(\tau)
$ and
$\left( \eta(\tau) \right)^{2/5} \,
  \Phi_2(\tau)
$
with weight $1/5$,
where we use
\begin{align}
  \boldsymbol{\chi}_{2,5}(\tau)
  & =
  \begin{pmatrix}
    \Phi_1(\tau)
    \\[2mm]
    \Phi_2(\tau)
  \end{pmatrix}
  =
  \begin{pmatrix}
    \displaystyle
    \frac{1}{\eta(\tau)} \sum_{n \in \mathbb{Z}}
    (-1)^n \, q^{\frac{1}{40} (10 n +1)^2}
    \\[4mm]
    \displaystyle
    \frac{1}{\eta(\tau)} \sum_{n \in \mathbb{Z}}
    (-1)^n \, q^{\frac{1}{40} (10 n +3)^2}
  \end{pmatrix}
  \\
  & = 
  \begin{pmatrix}
    q^{-\frac{1}{60}} \,
    \left(
      1 + q + q^2 + q^3 + \cdots
    \right)
    \\[2mm]
    q^{\frac{11}{60}} \,
    \left(
      1 + q^2 + q^3 + \cdots
    \right)
  \end{pmatrix}
  \nonumber
\end{align}
The  transformation laws of these $q$-series are given by
\begin{equation}
  \begin{gathered}
    \boldsymbol{\chi}_{2,5}(\tau)
    =
    \frac{2}{\sqrt{5}}  \,
    \begin{pmatrix}
      \sin \left( \frac{2}{5} \, \pi \right) &
      \sin \left( \frac{1}{5} \, \pi \right) 
      \\[2mm]
      \sin \left( \frac{1}{5} \, \pi \right) &
      - \sin \left( \frac{2}{5} \, \pi \right) 
    \end{pmatrix}
    \,
    \boldsymbol{\chi}_{2,5}(-1/\tau)
    \\[2mm]
    \boldsymbol{\chi}_{2,5}(\tau+1)
    =
    \begin{pmatrix}
      \E^{- \frac{1}{30} \pi \I} & 
      \\[2mm]
      & \E^{\frac{11}{30} \pi \I}
    \end{pmatrix}
    \,
    \boldsymbol{\chi}_{2,5}(\tau)
  \end{gathered}
\end{equation}
It should be noted that the rational weight plays a crucial role in studying
congruence subgroup in Ref.~\citen{TIbuki00a}, but here we choose
$\boldsymbol{\chi}_{2,5}(\tau)$
to be
weight-zero  modular form
from  the point of view of the
conformal field theory,
because
these are the Virasoro  characters of the minimal model
$\mathcal{M}(2,5)$, or the Lee--Yang theory~\cite{Rocha84a}.
They can be written as follows due to  the Rogers--Ramanujan
identity and the Jacobi triple product formula;
\begin{equation*}
  \begin{aligned}
    q^{ \frac{1}{60}} \,    \Phi_1 (\tau)
    & =
    \prod_{n=0}^\infty
    \frac{1}{
      \left( 1 -q^{5n+1} \right) \,
      \left( 1 -q^{5n+4} \right)
    }
    \\
    &    =
    \sum_{n=0}^\infty \frac{q^{n^2}}{
      \prod_{k=1}^n
      \left( 1-q^k \right)
    }
    \\[2mm]
    q^{- \frac{11}{60}} \, 
    \Phi_2 (\tau)
    & =
    \prod_{n=0}^\infty
    \frac{1}{
      \left( 1 -q^{5n+2} \right) \,
      \left( 1 -q^{5n+3} \right)
    }
    \\
    & =
    \sum_{n=0}^\infty
    \frac{q^{n^2+n}}{
      \prod_{k=1}^n
      \left( 1- q^k \right)
    }
  \end{aligned}
\end{equation*}
See Refs.~\citen{KHikami02c,KHikami03b} for recent studies on the
Rogers--Ramanujan type generating function of the $L$-function
as a generalization of Zagier's identity~\cite{DZagier92a,KOno04Book,AndrUrroOnok01a}.
We should remark  that the Eichler integral of the modular form
$\eta(\tau) \cdot \Phi_2(\tau)$ with weight $1/2$
coincides with Kashaev's
invariant for torus knot
$\mathcal{T}_{2,5}$~\cite{KHikami03c,KHikami02b}.

To see a  relationship between
these two bases, $\boldsymbol{\Psi}_{E_8}(\tau)$ and
$\boldsymbol{\chi}_{2,5}(\tau)$, of the group $\Gamma(5)$,
we recall that
invariant polynomials for the  vector modular form
$\boldsymbol{\chi}_{2,5}(\tau)$
have the same form
with~\eqref{icosahedron_Klein_polynomial} replacing $(X,Y)$ with
$(\Phi_2, \Phi_1)$.
Explicitly  they are computed as
\begin{equation}
  \label{E8_polynomial_Phi}
  \begin{gathered}
    V_I( \Phi_2, \Phi_1)
    =
    -1
    \\[2mm]
    \left( \eta(\tau) \right)^8 \cdot
    F_I( \Phi_2, \Phi_1)
    = E_4
    \\[2mm]
    \left( \eta(\tau) \right)^{12} \cdot
    E_I(  \Phi_2, \Phi_1)
    = E_6
  \end{gathered}
\end{equation}
Equating~\eqref{E8_polynomial} with~\eqref{E8_polynomial_Phi}, we find that
\begin{equation}
  \begin{aligned}
    X
    & =
    \Phi_1^{~2}
    \left(
      \Phi_1^{~5}
      + 7 \,
      \Phi_2^{~5}
    \right)
    \\[2mm]
    Y
    & =
    \left(
      7 \, \Phi_1^{~5}
      -
      \Phi_2^{~5}
    \right) \,
    \Phi_2^{~2}
  \end{aligned}
\end{equation}
which coincides with one of solutions in Ref.~\citen{MKanek04a}.
Therein
the Fuchsian differential equation~\cite{MatMukSen88a,EBKirit89a}
was investigated, and 
given explicitly were homogeneous polynomials of $\Phi_1$ and $\Phi_2$
which constitute  the two dimensional vector modular space.

To conclude
we have a mapping
\begin{equation*}
  \boldsymbol{\Psi}_{E_8}
  :
  \overline{\mathbb{H}/\Gamma(5)}
  \to \mathbb{P}^1
\end{equation*}
and the modular form $\boldsymbol{\Psi}_{E_8}$ is related to the
icosahedral group.

It may  help our understanding
to discuss a direct connection between
the homogeneous polynomials
$V_I$, $F_I$, and $E_I$~\eqref{icosahedron_Klein_polynomial}
and  the regular icosahedron.
We consider the unit sphere $S^2$ in $\mathbb{R}^3$, and inscribe the
regular icosahedron
in it with the north and south poles as two of
vertices thereof.
We identify $S^2$ with $\mathbb{P}^1$ as before,
and take the coordinates of $\mathbb{P}^1$ as $(X: Y)$.
Then the zeros of the polynomial $V_I$  coincides with the vertices of
the icosahedron
while the zeros of the polynomials $F_I$ and $E_I$ denote
the vertices of the dual  dodecahedron,
or the mid-points of face of the icosahedron,
and the mid-edge points respectively.
Therefore these polynomials are invariant under the icosahedral group.

\subsection{\mathversion{bold}
  $\mathcal{D}_3$}

A realization of the cube in $\mathbb{P}^1$
appears  in the modular forms for the manifold
$\mathcal{D}_3$.
As pointed out in~\eqref{D_odd_link}, the WRT invariant for the
manifold $\mathcal{D}_3$ is related to  Kashaev's invariant for torus
link $\mathcal{T}_{2,6}$.
{}From the viewpoint of  modular forms, these two quantum
invariants can be regarded as the Eichler integrals of the following
two-dimensional vector modular form;
\begin{align}
  \boldsymbol{\Psi}_{D_5}(\tau)
  & =
  \frac{1}{
    \left( \eta(\tau) \right)^3
    } \,
  \begin{pmatrix}
    \Psi_3^{(1)}(\tau) 
    \\[2mm]
    \Psi_3^{(2)}(\tau) 
  \end{pmatrix}
  =
  \begin{pmatrix}
    \displaystyle
    \frac{
      \left( \eta(2 \, \tau) \right)^5
    }{
      \left( \eta( \tau) \right)^3 \,
      \left( \eta(4 \, \tau) \right)^2
    }
    \\[6mm]
    \displaystyle
    2 \, 
    \frac{
      \left(  \eta(4 \, \tau) \right)^2
    }{
      \eta( \tau) \,
      \eta(2 \, \tau) 
    }
  \end{pmatrix}
  \\
  & \equiv \nonumber
  \begin{pmatrix}
    X(\tau)
    \\[2mm]
    Y(\tau)
  \end{pmatrix}
  =
  \begin{pmatrix}
    q^{-\frac{1}{24}} \, \left(
      1+ 3 \, q + 4 \, q^2 + 7 \, q^3 + \cdots
    \right)
    \\[4mm]
    2 \, q^{\frac{5}{24}} \, \left(
      1 + q + 3 \, q^2 + 4 \, q^3 + \cdots
    \right)
  \end{pmatrix}
\end{align}
This transforms as
\begin{equation}
  \label{ST_D5}
  \begin{gathered}
    \boldsymbol{\Psi}_{D_5}(\tau)
    =
    \frac{1}{\sqrt{2}} \,
    \begin{pmatrix}
      1 & 1 \\[2mm]
      1 & -1
    \end{pmatrix} \,
    \boldsymbol{\Psi}_{D_5}(-1/\tau)
    \\[2mm]
    \boldsymbol{\Psi}_{D_5}(\tau+1)
    =
    \begin{pmatrix}
      \E^{-\frac{1}{12} \pi \I} &  \\[2mm]
      & \E^{\frac{5}{12} \pi \I}
    \end{pmatrix} \,
    \boldsymbol{\Psi}_{D_5}(\tau)
  \end{gathered}
\end{equation}

The  homogeneous invariant polynomials of $X$ and $Y$ are then given
by~\cite{Klein56Book}
\begin{equation}
  \label{D5_EFV}
  \begin{gathered}
    F_C = X \, Y \, \left(
      X^4 - Y^4
    \right)
    \\[2mm]
    V_C =
    X^8 + 14 \, X^4 \, Y^4 + Y^8
    \\[2mm]
    E_C =
    X^{12} - 33 \, X^8 \, Y^4 - 33 \, X^4 \, Y^8 + Y^{12}
  \end{gathered}
\end{equation}
{}From the transformation law under the modular group,
we see that
these polynomials can be written in terms of the Eisenstein series as
\begin{equation}
  \begin{gathered}
    F_C(\tau) = 2 
    \\[2mm]
    \left( \eta(\tau)\right)^{8} \,
    V_C(\tau)
    =
    E_4(\tau)
    \\[2mm]
    \left( \eta(\tau) \right)^{12} \,
    E_C(\tau)
    = E_6(\tau)
  \end{gathered}
\end{equation}
By definition~\eqref{D5_EFV} the invariant polynomials
fulfill the cubic equation
\begin{equation}
  V_C^{~3} - 108 \, F_C^{~4} = E_C^{~2}
\end{equation}
If we set
$x= 12 \sqrt{3} \, E_C \, F_C^{~2}$,
$y=V_C^{~2}$,
and $z=\I \, \left(
  E_C^{~2}-108 \, F_C^{~4}
\right)$,
we recover
\begin{equation*}
  R(x,y,z)
  =
  x^2 \, y + y^4 + z^2=0
\end{equation*}

{}From the viewpoint of the principal congruence subgroup 
we  note that the vector modular form is written as
\begin{equation*}
  \eta(\tau) \,
  \boldsymbol{\Psi}_{D_5}(\tau)
  =
  \begin{pmatrix}
    \displaystyle
    \sum_{n \in 2 \mathbb{Z}} q^{\frac{1}{4} n^2}
    \\[8mm]
    \displaystyle
    \sum_{n \in 2 \mathbb{Z}+1} q^{\frac{1}{4} n^2}
  \end{pmatrix}
\end{equation*}
The first component  in the right hand side
denotes the theta function  on the root lattice $A_1$,
\emph{i.e.}, the lattice
$\sqrt{2} \, \mathbb{Z}$,
and the sum of two components
becomes a theta series  on the dual lattice
$\frac{1}{\sqrt{2}}\, \mathbb{Z}$.
As the level of  the lattice $\sqrt{2} \, \mathbb{Z}$ is 4, the right
hand side is  the modular form for the subgroup $\Gamma(4)$.
The group $PSL(2;\mathbb{Z})/\Gamma(4)$ is isomorphic to the cubic
group, and
we have a mapping
\begin{equation*}
  \boldsymbol{\Psi}_{D_5}
  :
  \overline{\mathbb{H}/\Gamma(4)}
  \rightarrow \mathbb{P}^1
\end{equation*}

This correspondence may be explained simply as follows~\cite{Ebeli94Book}.
We consider the unit sphere $S^2$ around the origin, and 
draw a cube
inscribed therein with faces perpendicular to the coordinate
axes.
As before, we  identify $S^2$ with $\mathbb{P}^1$, and  set $(X: Y)$
as the homogeneous coordinates of $\mathbb{P}^1$.
Then the zeros of the polynomial $V_C$ denote the eight vertices of
the cube.
Correspondingly the zeros of $F_C$ and $E_C$ are mid-points of
faces and edges
respectively, and it is natural that the  invariant polynomials have a
form of~\eqref{D5_EFV}.

\subsection{\mathversion{bold}
  $\mathcal{D}_2$}

As a final example, we briefly study the manifold $\mathcal{D}_2$.
The WRT invariant for this manifold is the Eichler integral of the
modular form with weight $3/2$
\begin{equation}
  \Psi_2^{(1)}(\tau)
  =
  \left(
    \eta(\tau)
  \right)^3
\end{equation}
In terms of the Jacobi theta function~\eqref{Jacobi_theta}, this
modular form can be factorized as
\begin{equation}
  2 \, 
  \left(
    \eta(\tau)
  \right)^3
  =
  \theta_{00}(\tau) \,
  \theta_{01}(\tau) \,
  \theta_{10}(\tau) 
\end{equation}
These theta functions satisfy
\begin{equation}
  \left( \theta_{10}(\tau) \right)^{4} +
  \left( \theta_{01}(\tau) \right)^{4} =
  \left( \theta_{00}(\tau) \right)^{4}
\end{equation}
This algebraic equation may be identified with that in
Table~\ref{tab:surface};
\begin{equation*}
  R(x,y,z)
  =
  x^2 \, y + y^3 + z^2 = 0
\end{equation*}
after setting
$x=- \I \,
\left(
  \theta_{01}^{~4} -   \theta_{10}^{~4}
\right)$,
$y = - \theta_{00}^{~4}$,
and $z=2 \,
\left(
  \theta_{00} \, \theta_{01} \, \theta_{10}
\right)^2$.

\section{Conclusions and Discussions}

We have revealed the connection between the SU(2) WRT invariants and
modular forms.
We have shown that the  WRT invariant for the spherical Seifert manifolds
$S^3/\Gamma$ with a finite subgroup $\Gamma$ can be written in terms
of the Eichler integrals of  modular forms with weight $3/2$.
Explicit forms are
given in
\eqref{WRT_E6},~\eqref{WRT_E7},~\eqref{WRT_E8},~\eqref{WRT_Deven},
and~\eqref{WRT_Dodd}, and they
suggest that the WRT invariants may be decomposed by the  torsion
linking pairing,
$\lambda: \Tors H_1(\mathcal{M};\mathbb{Z}) \otimes
\Tors H_1(\mathcal{M};\mathbb{Z}) \to \mathbb{Q}/\mathbb{Z}
$, as~\footnote{This observation is due to K.~Habiro. The author
  thanks him for pointing out.}
\begin{equation*}
  \tau_N(\mathcal{M})
  =
  \sum_\lambda \E^{2 \pi \I  \lambda  N} \,
  \tau_N^{(\lambda)}(\mathcal{M})
\end{equation*}
where $\tau_N^{(\lambda)}(\mathcal{M})$ is a limiting value of holomorphic
$q$-series at the $N$-th root of unity.
We can check from a result of Ref.~\citen{LCJeff92a}
that
this decomposition is fulfilled for lens space.

Based on the nearly modular property of the Eichler integrals, we have
obtained the exact asymptotic expansion of the WRT invariants in
$N\to\infty$.
We have checked that a dominating term of the Witten  partition function
$Z_{N-2}(\mathcal{M})$ can
be written in terms of the classical topological invariants
as~\eqref{Z_asymptotics}.
Our results are  summarized  in Table~\ref{tab:CS}.
We see that an inverse of the Euler characteristic coincides with
subscript $P$ of the Eichler integrals
$\widetilde{\Psi}_P^{(a)}(1/N)$.
As seen from the nearly modular transformation
formula~\eqref{Psi_nearly_modular}, the Chern--Simons invariant is
related to  an exponential factor of 
a limiting value of the Eichler integrals at integers
$\widetilde{\Psi}_P^{(a)}(-N)$, while
both
the absolute value  of the Eichler integrals at integers
and matrix elements of the modular $S$-matrix 
are related to the torsion.

\begin{table}[htbp]
  \centering
  \begin{equation*}
    \renewcommand{\arraystretch}{1.6}
    \begin{array}{c||c|c|c|}
      \mathcal{M} & e(\mathcal{M}) & \text{Eichler integrals} &
      \text{Chern--Simons invariant $\CS(A_\alpha)$}
      \\
      \hline \hline
      M(2,3,3) & \frac{1}{6} & \widetilde{\Psi}_6^{(1), (3), (5)} &
      \left\{
        -\frac{1}{24}
      \right\}
      \\
      M(2,3,4) & \frac{1}{12} & \widetilde{\Psi}_{12}^{(1), (5), (7), (11)} &
      \left\{
        -\frac{1}{48} ,         -\frac{25}{48}
      \right\}
      \\
      M(2,3,5) & \frac{1}{30} & \widetilde{\Psi}_{30}^{(1), (11),
        (19), (29)} &
      \left\{
        -\frac{1}{120},         -\frac{49}{120}
      \right\}
      \\
      M(2,2,K) & \frac{1}{K} & \widetilde{\Psi}_{K}^{(1), (K-1)} &
      \text{
        $      \left\{
          -\frac{(2 \, m +1)^2}{4 \, K}
        \right\}$
        for $0 \leq m < \frac{K-1}{2}$
      }
      \\
      \hline
    \end{array}
  \end{equation*}
  \caption{Relationship between the WRT invariants and the Eichler
    integrals is given.
    The Eichler integrals $\widetilde{\Psi}_P^{(a), (b), \dots}$ means
    that the WRT invariant $\tau_N(\mathcal{M})$ is
    written as a linear combination of the Eichler integrals
    $\widetilde{\Psi}_P^{(a)}(1/N)$,
    $\widetilde{\Psi}_P^{(b)}(1/N)$, $\cdots$.
  }
  \label{tab:CS}
\end{table}

Moreover we have  clarified that  the modular forms,
whose Eichler
integrals
contribute to the quantum invariants,
have connections
with the polyhedral group.
We have studied the invariant polynomials of the modular group, and 
we have found that
they construct the polyhedral equations
(see Tables~\ref{tab:spherical} and~\ref{tab:surface}).
Pointed out
in Ref.~\citen{GuadPilo98a} is that the absolute value of the WRT
invariant depends on the fundamental group.
Our results prove that the WRT invariant has some informations about the
fundamental group of  manifolds.
As   the WRT invariant for the Seifert homology
spheres 
can be written in terms of the Eichler integrals of 
half-integral weight modular
forms as was studied in Refs.~\citen{KHikami04b,KHikami04e,KHikami04f},
studies on geometry of modular forms will bring us  fruitful
insights on   geometry of the
quantum invariants
even though the fundamental group is no longer finite.


We take 
the Brieskorn homology sphere
$\Sigma(2,3,7)$
as  an example.
The fundamental group is not finite, and it corresponds to a
hyperbolic tessellation~\cite{JMiln75a}.
The WRT invariant for $\Sigma(2,3,7)$ is identified with
$\widetilde{\Phi}_{2,3,7}^{(1,1,1)}(1/N)$ which is the Eichler
integral of  modular form with weight $3/2$~\cite{KHikami04b}.
This modular form spans a 3-dimensional space  with two more
$q$-series;
we  introduce the  vector modular form
$\boldsymbol{\Phi}_{2,3,7}(\tau)$ by
\begin{align}
  \boldsymbol{\Phi}_{2,3,7}(\tau)
  & =
  \frac{1}{
    \left(
      \eta(\tau)
    \right)^3
  }  \,
  \begin{pmatrix}
    \Phi_{2,3,7}^{(1,1,1)}(\tau)
    \\[2mm]
    \Phi_{2,3,7}^{(1,1,2)}(\tau)
    \\[2mm]
    \Phi_{2,3,7}^{(1,1,3)}(\tau)
  \end{pmatrix}
  \\
  & \equiv \nonumber
  \begin{pmatrix}
    X(\tau) \\[2mm]
    - Y(\tau)  \\[2mm]
    - Z(\tau)
  \end{pmatrix}
  =
  \begin{pmatrix}
    q^{- \frac{5}{42}} \,
    \left(
      1 - 10 \, q -30 \, q^2 - 95 \, q^{3}
      - \cdots
    \right)
    \\[4mm]
    - q^{\frac{1}{42}} \,
    \left(
      5 + 15\, q + 64 \, q^2 + 190 \, q^3 +
      \cdots
    \right)
    \\[4mm]
    - q^{\frac{25}{42}} \,
    \left(
      11+ 50 \, q + 150 \, q^2 + 420 \, q^3
      + \cdots
    \right)
  \end{pmatrix}
\end{align}
where each element is defined by
\begin{equation*}
  \Phi_{2,3,7}^{\vv{\ell}}(\tau)
  =
  \frac{1}{2} \sum_{n \in \mathbb{Z}} n \, \chi_{84}^{\vv{\ell}}(n) \,
  q^{\frac{n^2}{168}}
\end{equation*}
for a triple $\vv{\ell}= (\ell_1, \ell_2, \ell_3)$,
and  $\chi_{84}^{\vv{\ell}}(n)$ is the odd periodic function with
modulus 84;
\begin{align*}
  \chi_{84}^{(1,1,1)}(n)
  & =
  \psi_{84}^{(1)}(n)  -   \psi_{84}^{(13)}(n) 
  -   \psi_{84}^{(29)}(n)  +   \psi_{84}^{(41)}(n) 
  \\[2mm]
  \chi_{84}^{(1,1,2)}(n)
  & =
  - \psi_{84}^{(5)}(n)  -   \psi_{84}^{(19)}(n) 
  -   \psi_{84}^{(23)}(n)  -   \psi_{84}^{(37)}(n) 
  \\[2mm]
  \chi_{84}^{(1,1,3)}(n)
  & =
  - \psi_{84}^{(11)}(n)  -   \psi_{84}^{(17)}(n) 
  -   \psi_{84}^{(25)}(n)  -   \psi_{84}^{(31)}(n) 
\end{align*}
The transformation properties are given by~\cite{KHikami04b}
\begin{equation}
  \label{237_modular_transformation}
  \begin{gathered}
    \boldsymbol{\Phi}_{2,3,7}(\tau)
    =
    \frac{-2}{\sqrt{7}} \,
    \begin{pmatrix}
      \sin \left( \frac{\pi}{7} \right)
      & \sin \left(\frac{2 \, \pi}{7}  \right)
      &  \sin \left(\frac{3 \, \pi}{7} \right)
      \\[2mm]
      \sin \left(\frac{2 \,\pi}{7} \right)
      &    -   \sin\left( \frac{3 \, \pi}{7}\right)
      &    \sin \left(\frac{ \pi}{7} \right)
      \\[2mm]
      \sin \left( \frac{3 \, \pi}{7} \right)
      &       \sin \left( \frac{\pi}{7} \right)
      &  -\sin \left(\frac{2 \, \pi}{7} \right)
    \end{pmatrix}
    \,
    \boldsymbol{\Phi}_{2,3,7}(-1/\tau)
    \\[2mm]
    \boldsymbol{\Phi}_{2,3,7}(\tau+1)
    =
    \begin{pmatrix}
      \E^{ -\frac{5}{21} \pi \I} &&
      \\
      &
      \E^{ \frac{1}{21} \pi \I}
      \\
      &&
      \E^{  \frac{25}{21} \pi \I}
    \end{pmatrix}
    \,
    \boldsymbol{\Phi}_{2,3,7}(\tau)
  \end{gathered}
\end{equation}
Due to that  coefficients of $q$-series
$X(\tau)$  have both positive and negative integers
(constant term is $+1$ while
coefficients of
positive powers of $q$ are negative),
we are not sure whether
this vector modular form is related to  character of the conformal field
theory as in the case of the Poincar{\'e} homology sphere.

Following results
on the  Klein quartic~\cite{Klein79a}, we define the
homogeneous polynomials $F_Q$, $G_Q$,  and $H_Q$ by
\begin{equation}
  \label{237_polynomial}
  \begin{gathered}
    F_Q(X,Y,Z)
    = X^3 \, Y + Y^3 \, Z + Z^3 \, X
    \\[2mm]
    G_Q(X,Y,Z)
    = X \, Y^5 + Y \, Z^5 + Z \, X^5 - 5 \, X^2 \, Y^2 \, Z^2
    \\[2mm]
    \begin{aligned}
      H_Q(X,Y,Z)
      & =
      X^{14} + Y^{14} + Z^{14}
      - 34 \,
      \left(
        X^{11} \, Y^2 \, Z +  X^{2} \, Y \, Z^{11} +  X \, Y^{11} \, Z^2
      \right)
      \\
      -  
      250 \, &
      \left(
        X^9 \, Y \, Z^4 + X \, Y^4 \, Z^9 + X^4 \, Y^9 \, Z
      \right)
      + 375 \,
      \left(
        X^8 \, Y^4 \, Z^2 + X^4 \, Y^2 \, Z^8 + X^2 \, Y^8 \, Z^4
      \right)
      \\
      &
      + 18 \,
      \left(
        X^7 \, Y^7 + Y^7 \, Z^7 + Z^7 \, X^7
      \right)
      - 126 \,
      \left(
        X^6 \, Y^3 \, Z^5 + X^3 \, Y^5 \, Z^6 + X^5 \, Y^6 \, Z^3
      \right)
    \end{aligned}
  \end{gathered}
\end{equation}
We can check that these are invariant polynomials
under~\eqref{237_modular_transformation}, and  by use of the
$q$-series expansion we find
that they are given in terms of the Eisenstein series and the Dedekind
$\eta$-function as
\begin{equation}
  \begin{gathered}
    \left( \eta(\tau) \right)^8 \, F_Q
    =
    5 \, E_4(\tau)
    \\[2mm]
    G_Q =
    3136 
    \\[2mm]
    \left(\eta(\tau)\right)^{40} \,
    H_Q
    =
    \frac{1}{27} \,
    (E_4)^{2} \,
    \left(
      21832 \, (E_4)^3 - 21805 \, (E_6)^2
    \right)
  \end{gathered}
\end{equation}
As a consequence of~\eqref{E_E_Delta}, we obtain an algebraic
equation of the invariant polynomials as
\begin{equation}
  G_Q \, H_Q
  =
  \frac{3136}{3125} \, F_Q^{~5} + 
  \frac{89}{5} \, F_Q^{~2} \, G_Q^{~2}
\end{equation}

The modular transformation law~\eqref{237_modular_transformation}
shows that the  modular form 
$\left(
\eta(\tau)
\right)^{20/7} \, \boldsymbol{\Psi}_{2,3,7}(\tau)$
with rational weight $10/7$
is on the group
$\Gamma(7)$.
Previously known
basis of polynomial ring associated to the group
$\Gamma(7)$ is the modular form
$\left(
  \eta(\tau)
\right)^{4/7} \,
\boldsymbol{\chi}_{2,7}(\tau)$
with weight $2/7$
where we mean~\cite{TIbuki00a}
\begin{align}
  \label{chi_27_xyz}
  \boldsymbol{\chi}_{2,7}(\tau)
  & =
  \begin{pmatrix}
    - x (\tau)
    \\[2mm]
    y (\tau)
    \\[2mm]
    z (\tau)
  \end{pmatrix}
  =
  \frac{1}{\eta(\tau)} \,
  \begin{pmatrix}
    \displaystyle
    q^{\frac{25}{56}} \,
    \prod_{n=1}^\infty
    \left( 1 - q^{7 n} \right) \,
    \left( 1 - q^{7 n-6} \right) \,
    \left( 1 - q^{7 n-1} \right)
    \\[4mm]
    \displaystyle
    q^{\frac{9}{56}} \,
    \prod_{n=1}^\infty
    \left( 1 - q^{7 n} \right) \,
    \left( 1 - q^{7 n-5} \right) \,
    \left( 1 - q^{7 n-2} \right)
    \\
    \displaystyle
    q^{\frac{1}{56}} \,
    \prod_{n=1}^\infty
    \left( 1 - q^{7 n} \right) \,
    \left( 1 - q^{7 n-4} \right) \,
    \left( 1 - q^{7 n-3} \right)
  \end{pmatrix}
  \\
  & =
  \begin{pmatrix}
    q^{\frac{17}{42}} \,
    \left(
      1 + q^2 + q^3+ \cdots
    \right)
    \\[2mm]
    q^{\frac{5}{42}} \,
    \left(
      1 + q + q^2 + 2 \, q^3 + \cdots
    \right)
    \\[2mm]
    q^{-\frac{1}{42}} \,
    \left(
      1 + q + 2 \, q^2 + 2 \, q^3 + \cdots
    \right)
  \end{pmatrix}
  \nonumber
\end{align}
It should be noted that  the weight-zero vector
$\boldsymbol{\chi}_{2,7}(\tau)$
coincides
with the character of the Virasoro minimal model
$\mathcal{M}(2,7)$~\cite{Rocha84a}.
In general the theta function
basis in Ref.~\citen{TIbuki00a} is the
character of the Virasoro minimal model $\mathcal{M}(2,N)$ for odd
$N$ up to fractional powers of the Dedekind $\eta$-function,
and as was shown in Refs.~\citen{KHikami02b,KHikami03c}
their  Eichler integrals  are   proportional to Kashaev's invariant for
torus knot $\mathcal{T}_{2,N}$.
The modular transformation of $\boldsymbol{\chi}_{2,7}(\tau)$ is
given as
\begin{equation}
  \begin{gathered}
    \boldsymbol{\chi}_{2,7}(\tau)
    =
    \frac{2}{\sqrt{7}} \,
    \begin{pmatrix}
      \sin \left( \frac{2 \, \pi}{7} \right)
      &  - \sin \left(\frac{3 \, \pi}{7}  \right)
      &  \sin \left(\frac{ \pi}{7} \right)
      \\[2mm]
      - \sin \left(\frac{3 \,\pi}{7} \right)
      &    -   \sin\left( \frac{ \pi}{7}\right)
      &    \sin \left(\frac{2\, \pi}{7} \right)
      \\[2mm]
      \sin \left( \frac{ \pi}{7} \right)
      &       \sin \left( \frac{2\, \pi}{7} \right)
      &  \sin \left(\frac{3 \, \pi}{7} \right)
    \end{pmatrix}
    \,
    \boldsymbol{\chi}_{2,7}(-1/\tau)
    \\[2mm]
    \boldsymbol{\chi}_{2,7}(\tau+1)
    =
    \begin{pmatrix}
      \E^{ \frac{17}{21} \pi \I} &&
      \\
      &
      \E^{ \frac{5}{21} \pi \I}
      \\
      &&
      \E^{ - \frac{1}{21} \pi \I}
    \end{pmatrix} \,
    \boldsymbol{\chi}_{2,7}(\tau)
  \end{gathered}
\end{equation}
The invariant polynomials under these transformations
have the same form with~\eqref{237_polynomial} and~\eqref{237_W_polynomial}
replacing $(X,Y,Z)$ with $(x,y,z)$ in~\eqref{chi_27_xyz},
and  by simple computations we obtain~\cite{NElki99a}
\begin{equation}
  \label{base_xyz}
  \begin{gathered}
    F_Q(x,y,z)=0
    \\[2mm]
    G_Q(x,y,z)=1
    \\[2mm]
    \left( \eta(\tau) \right)^8 \,
    H_Q(x,y,z)
    =
    E_4(\tau)
    \\[2mm]
    \left( \eta(\tau) \right)^{12} \,
    W_Q(x,y,z)
    =
    E_6(\tau)
  \end{gathered}
\end{equation}
where we have used 
one more invariant polynomial of order 21
defined by the Jacobian
\begin{equation}
  \label{237_W_polynomial}
  W_Q
  (x, y, z)
  =
  \frac{1}{14} \, \frac{
    \partial \, \left(
      F_Q, G_Q, H_Q
    \right)
  }{
    \partial
    \left(
      x, y, z
    \right)
  }
\end{equation}
We have an algebraic relation between these 4 invariant
polynomials as
\begin{multline}
  \label{algebraic_237}
  W_Q^{~2} = H_Q^{~3} - 1728 \, G_Q^{~7}
  + 1008 \, F_Q \, G_Q^{~4} \, H_Q - 32 \, F_Q^{~2} \, G_Q \, H_Q^{~2}
  + 19712 \, F_Q^{~3} \, G_Q^{~5}
  \\
  - 1152 \, F_Q^{~4} \, G_Q^{~2} \, H_Q
  + 11264  \, F_Q^{~6} \, G_Q^{~3}
  - 256 \, F_Q^{~7} \, H_Q
  + 12288 \, F_Q^{~9} \, G_Q
\end{multline}

Result of Ref.~\citen{TIbuki00a} shows that the polynomial ring of
$(x,y,z)$  is on $\Gamma(7)$, and
we find that our basis $(X,Y,Z)$ is  in fact given by
\begin{equation}
  \begin{aligned}
    X & =
    z^5 - 10 \, x^2 \, y \, z^2 + 5 \, x \, y^4
    \\[2mm]
    Y & =
    x^5 - 10 \, x^2 \, y^2 \, z + 5 \, y \, z^4
    \\[2mm]
    Z & =
    y^5 - 10 \, x \, y^2 \, z^2 + 5 \, x^4 \, z
  \end{aligned}
\end{equation}
It is interesting  to study, as a generalization of Ref.~\citen{MKanek04a},
the third-order Fuchsian differential
equation, and to find the homogeneous polynomials which constitute the
three dimensional vector modular space.

As a result,
the modular form $\boldsymbol{\Phi}_{2,3,7}(\tau)$ is  on
$\Gamma(7)$, and we have a mapping
\begin{equation*}
  \boldsymbol{\Phi}_{2,3,7} :
  \overline{\mathbb{H}/\Gamma(7)}
  \rightarrow
  \mathbb{P}^2
\end{equation*}
Furthermore 
in the basis of $(x,y,z)$
an algebraic equation~\eqref{algebraic_237} reduces to 
\begin{equation}
  \label{relation_237}
  W_Q^{~2} + 1728 \,  G_Q^{~7} = H_Q^{~3}
\end{equation}
due to  a condition of the Klein quartic $F_Q=0$~\eqref{base_xyz}.
This equation  has the $E_{12}$-type  
exceptional singularity of Arnold,
and it is obtained  by hyperbolic tessellation of a triangle
$\left(
\frac{\pi}{2},
\frac{\pi}{3},
\frac{\pi}{7}
\right)$.
The algebraic equation~\eqref{relation_237} should be compared
with~\eqref{Brieskorn_surface}, and
the fundamental group of
the Brieskorn sphere $\Sigma(2,3,7)$ indeed denotes the reflection
group of this hyperbolic triangle.

\section*{Acknowledgments}
The author would like to thank T.~Shioda who raised a question about
the symmetry  of modular form $\boldsymbol{\Psi}_{E_8}(\tau)$.
Thanks are also to T.~Takata for bringing
Refs.~\citen{GuadPilo98a,SYamad95a} to attention.
The author would like to thank 
A. N. Kirillov and H.~Murakami
for private communications.
This work is supported in part  by Grant-in-Aid for Young Scientists
from the Ministry of Education, Culture, Sports, Science and
Technology of Japan.


\end{document}